\documentclass[12pt, preprint]{aastex}
\usepackage{epsfig}
\usepackage{natbib}
\usepackage{amsmath}
\shorttitle{Multiplicity of Massive Stars}
\shortauthors{Aldoretta et al.}

\newcommand{\noprint}[1]{}
\newcommand{\figsetstart}{{\bf Fig. Set} }
\newcommand{\figsetend}{}
\newcommand{\figsetgrpstart}{}
\newcommand{\figsetgrpend}{}
\newcommand{\figsetnum}[1]{{\bf #1.}}
\newcommand{\figsettitle}[1]{ {\bf #1} }
\newcommand{\figsetgrpnum}[1]{\noprint{#1}}
\newcommand{\figsetgrptitle}[1]{\noprint{#1}}
\newcommand{\figsetplot}[1]{\noprint{#1}}
\newcommand{\figsetgrpnote}[1]{\noprint{#1}}

\begin{document}

\received{2014 May 30}
\setcounter{footnote}{0}

\title{The Multiplicity of Massive Stars: 
A High Angular Resolution Survey with the HST Fine Guidance 
Sensor\footnote{Based on observations made with the NASA/ESA Hubble Space Telescope, 
obtained at the Space Telescope Science Institute, which is operated by the 
Association of Universities for Research in Astronomy, Inc., under NASA 
contract NAS 5-26555. These observations are associated with programs  
11212, 11901, 11943, and 11944.}}

\author{
E. J. Aldoretta\altaffilmark{2,3},
S. M. Caballero-Nieves\altaffilmark{4},
D. R. Gies\altaffilmark{2},
E. P. Nelan\altaffilmark{5},
D. J. Wallace\altaffilmark{6,7},\\
W. I. Hartkopf\altaffilmark{8}, 
T. J. Henry\altaffilmark{2},
W.-C. Jao\altaffilmark{2},
J. Ma\'{i}z Apell\'{a}niz\altaffilmark{9},
B. D. Mason\altaffilmark{8},\\
A. F. J. Moffat\altaffilmark{3},
R. P. Norris\altaffilmark{2},
N. D. Richardson\altaffilmark{3}, and 
S. J. Williams\altaffilmark{10}
}

\email{
emily@astro.umontreal.ca,
s.caballero@sheffield.ac.uk,
gies@chara.gsu.edu,
nelan@stsci.edu,
debra.j.wallace@nasa.gov,
william.hartkopf@usno.navy.mil,
thenry@chara.gsu.edu,
jao@chara.gsu.edu,
jmaiz@iaa.es,
brian.mason@usno.navy.mil,
moffat@astro.umontreal.ca,
norris@chara.gsu.edu,
richardson@astro.umontreal.ca,
williams@astro.noa.gr
}

\altaffiltext{2}{Center for High Angular Resolution Astronomy, 
Department of Physics and Astronomy, 
Georgia State University, P.\ O.\ Box 5060,
Atlanta, GA 30302-5060, USA}
\altaffiltext{3}{D\'epartement de physique and Centre de
Recherche en Astrophysique du Qu\'ebec (CRAQ), Universit\'e de Montreal,
CP 6128 Succ. A., Centre-Ville, Montr\'eal, Qu\'ebec H3C 3J7, Canada}
\altaffiltext{4}{Department of Physics and Astronomy, University of Sheffield, 
Hicks Building, Hounsfield Road, Sheffield, S3 7RH, United Kingdom}
\altaffiltext{5}{Space Telescope Science Institute, 3700 San Martin Drive,
Baltimore, MD 21218, USA}
\altaffiltext{6}{Department of Natural Science, University of South Carolina -- Beaufort, 
801 Carteret Street, Beaufort, SC 29902, USA}
\altaffiltext{7}{Current address: NASA, Astrophysics Division, 
300 E Street, SW, Washington, DC 20546-0001, USA}
\altaffiltext{8}{US Naval Observatory, Astrometry Department, 3450 Massachusetts Avenue NW, 
Washington, DC 20392-5420, USA} 
\altaffiltext{9}{Instituto de Astrof\'{i}sica de Andaluc\'{i}a -- CSIC,
Glorieta de la Astronom\'{i}a, s/n.\ E-18008, Granada, Spain}
\altaffiltext{10}{Institute for Astronomy, Astrophysics, Space Applications and Remote Sensing (IAASARS), National Observatory of Athens, I. Metaxa \& Vas. Pavlou Street,
Palea Penteli, 15236 Athens, Greece}

\slugcomment{2014 July 28; submitted to AJ} 
\paperid{AJ-12127}


\begin{abstract}
We present the results of an all-sky survey made with the Fine Guidance Sensor 
on {\it Hubble Space Telescope} to search for angularly resolved binary 
systems among the massive stars.  The sample of 224 stars is comprised mainly
of Galactic O- and B-type stars and Luminous Blue Variables, plus a few 
luminous stars in the Large Magellanic Cloud.  
The FGS TRANS mode observations are sensitive 
to detection of companions with an angular separation between 
$0\farcs01$ and $1\farcs0$ and brighter than $\triangle m = 5$.  
The FGS observations resolved 52 binary and 6 triple star systems and 
detected partially resolved binaries in 7 additional targets 
(43 of these are new detections).  These numbers yield a companion detection 
frequency of $29\%$ for the FGS survey.  We also gathered 
literature results on the numbers of close spectroscopic binaries and 
wider astrometric binaries among the sample, and we present estimates of the 
frequency of multiple systems and the companion frequency for subsets of 
stars residing in clusters and associations, field stars, and runaway stars.
These results confirm the high multiplicity fraction, especially among 
massive stars in clusters and associations.  We show that the period 
distribution is approximately flat in increments of $\log P$. 
We identify a number of systems of potential interest for long term 
orbital determinations, and we note the importance of some of these 
companions for the interpretation of the radial velocities and light curves
of close binaries that have third companions. 
\end{abstract}

\keywords{binaries: general ---
stars: early-type ---
stars: massive ---
techniques: high angular resolution}

\setcounter{footnote}{0}


\section{Introduction}                                           

The formation of a star from a huge natal cloud presents a 
formidable problem of angular momentum redistribution 
\citep{2010RPPh...73a4901L}. 
Low mass stars may accomplish the removal of angular momentum through 
mass loss coupled with the stellar magnetic field 
\citep{2008ApJ...681..391M}. 
However, the situation appears to be different for the formation 
of massive stars that lack pervasive magnetic fields 
\citep{2007ARA&A..45..481Z}. 
These stars may form through disk accretion processes 
\citep{2009Sci...323..754K,2013ApJ...772...61K} 
and/or competitive accretion of smaller protostars
\citep{2005MNRAS.362..915B,2011IAUS..270...57B}; 
in both cases the angular momentum may be deposited into  
the orbital motion of nearby companion stars 
\citep{2008ApJ...681..375K,2012MNRAS.419.3115B}. 
Once formed, the massive binary systems may stand the best chance 
to survive the many dynamical encounters that probably occur
in dense cluster environments \citep{2011A&A...528A.144K}. 

There is ample evidence that the binary and multiple star frequency is 
remarkably high among massive stars \citep{2013ARA&A..51..269D}. 
Spectroscopic surveys of Galactic OB stars by \citet{2012MNRAS.424.1925C},
\citet{2012ApJ...756...50K}, \citet{2012Sci...337..444S}, and \citet{2014ApJS..211...10S},
and of the LMC Tarantula Nebula region by \citet{2013A&A...550A.107S}
demonstrate that the binary frequency may be $\approx 70\%$
for binaries with periods smaller than 3000~d.  
The incidence of longer period binaries has been explored through 
speckle interferometry by \citet{1998AJ....115..821M,2009AJ....137.3358M}, 
adaptive optics by \citet{2008AJ....136..554T} and \citet{2012ApJ...749..180C}, and 
Lucky Imaging by \citet{2010A&A...518A...1M} and \cite{2012A&A...538A..74P}. 
These studies also demonstrate the high incidence of binaries and 
multiples among longer period systems.  However, because of the 
great distances of most massive stars, there still exists a 
significant observational gap in our knowledge of binaries with 
periods of years to centuries that have radial velocity variations 
that are too small to measure or angular separations that are 
only resolvable with optical long baseline interferometry 
\citep{2009A&A...497..195K,2011AJ....142...21T,2011ApJ...740L..43S,2014Sana}.
It is critical to fill in this gap with new observations in 
order to determine the nature of the period distribution and 
to estimate the total binary frequency \citep{2013A&A...550A.107S}. 

The Fine Guidance Sensor (FGS) on the {\it Hubble Space Telescope} offers us 
a particularly attractive means to resolve such close visual binaries 
for even relatively faint targets \citep{2014fgsi.book.....N}. 
The prime FGS1r instrument is capable of resolving binaries as close as 
10 milliarcsec (mas) for stars as faint as $V=16$ mag.  
The FGS instrument was used to explore the binary frequency of massive
stars in two Galactic environments of special interest, the 
Carina Association by \citet{2004AJ....128..323N,2010AJ....139.2714N} and 
the Cyg OB2 association by \citet{2014AJ....147...40C}.  
In both cases the binary frequency over the angular range of 
$0\farcs01$ to $1\farcs0$ was found to be $\approx 22\%$.  

Here we describe a new all-sky FGS survey of the massive stars that we 
have made with a number of broad goals in mind.   Our primary task is 
to explore how binary properties vary with environment, in particular 
to search for evidence of different binary frequencies among massive stars
in clusters and associations and those in the field (especially runaway stars). 
A second goal is to compare the binary statistics in this angular range 
with those for close spectroscopic and wider separated systems in order to 
place constraints on the overall period distribution of massive binaries. 
Thirdly, we identify individual systems of particular interest where the 
distant companion may influence our interpretation of the spectra or 
light curve of the primary target and may serve for future mass determination 
through measured orbital motion.  We describe the observations and sample in \S2,
and then discuss the binary detection methods and results in \S3.  
The issues surrounding the companion frequency and period distribution 
are outlined in \S4 and \S5, respectively, and we summarize our conclusions in \S6. 

\setcounter{footnote}{10}

\section{FGS Observations}                                       

The Fine Guidance Sensor aboard the {\it Hubble Space Telescope} acts as
a single aperture, shearing interferometer that forms interference fringes
through a Koesters prism due to tilt differences in the incoming wavefront
\citep{2014fgsi.book.....N}.  There are three FGS instruments on {\it HST} that 
are used for precise pointing of the telescope, and one of these, FGS1r, is 
designated for science applications.  In the TRANS mode of operation the 
FGS1r scans across the target in two orthogonal directions, and it produces 
an $x$ and $y$ coordinate, fringe visibility curve (or ``S-curve'').
FGS observations of binary stars produce an S-curve that is the sum of 
fringe patterns for each component at a position that corresponds to 
the projection of the binary separation along the $x$ and $y$ vectors (\S3). 

The observations began as a SNAP program in Cycle 16 (GO-11212), and we 
selected targets all around the sky so that they could be easily scheduled 
into one orbit slots between other programs.  It was subsequently expanded 
into a Director's Discretionary program (GO-11901, 11943, 11944) around
the time of the last servicing mission in order to optimize telescope
usage when options with other instruments were very limited.   Consequently, 
the observations were made over the period 2007 to 2009 in a large number 
of single orbit pointings.  All the observations were made with the ND5 filter 
(brighter targets) or F583W filter (fainter targets) that record a broad 
range of the optical spectrum\footnote{http://www.stsci.edu/hst/fgs/design/filters}
($\approx 4600 - 7000$ \AA ). 
Multiple scans were recorded of each target with an angular step size of 1 mas, 
and the scans usually extended $\pm 1\farcs0$ from the main target 
(or longer in some cases where a wider companion was known).  
Note that the FGS detectors record all the flux from sources within 
the field of view (FOV $\approx 5\times 5$ arcsec$^2$), and although 
the detector response is uniform close to the target, the photometric response 
varies significantly for sources near the edge of the FOV.  Special 
calibration is necessary to obtain reliable magnitude differences for 
companions near the periphery.

All the observations were processed with the FGS pipeline software
\citep{2014fgsi.book.....N}.  First, the archived observations were 
extracted into individual scans using CALFGSA, which was also used 
to assess the data quality and to create a number of associated files
that document the properties of the scans and observations. 
Then we used the routine PTRANS to gather the individual scans, 
coalign them, and spatially smooth the combined results.   
Finally, we applied a simple spline fit to rectify the distant 
parts of each summed scan to a zero average.   

We selected our targets primarily from the Galactic O-Star 
Catalog\footnote{http://ssg.iaa.es/en/content/galactic-o-star-catalog/}
\citep{2004ApJS..151..103M,2013msao.confE.198M}, 
which we supplemented with other fainter targets from the catalog of
\citet{1974RMxAA...1..211C} 
and with a selection of Luminous Blue Variable (LBV) stars
\citep{2001A&A...366..508V}. 
Two interlopers were accidentally included in the sample, 
the hot subdwarf CD-45$^\circ$5058 = KS~292 
\citep{1991AnA...241..457R}
and the K-giant BD-3$^\circ$2178 
\citep{2010PASP..122.1437P}, which has been confused in 
the literature with the nearby hot subdwarf BD-3$^\circ$2179.
Both of these (apparently single) stars are excluded from the 
discussion in \S4 and \S5.  
The targets are listed in order of increasing right ascension in 
Table~1 , which provides the celestial coordinates, star name, 
the Johnson $V$ magnitude, and $B-V$ color \citep{1994cmud.book.....M}. 
Column 5 gives the spectral classification of the brightest component 
from (in most cases) \citet{2011ApJS..193...24S,2014ApJS..211...10S};
LBV classifications are from contemporaneous spectra described 
by \citet{2012ASPC..465..160R}.  
Columns 6 to 10 give information about the star's environment, 
runaway status, distance, spectroscopic binary status, and 
a recent spectroscopic reference, all gathered from a 
literature search for each object (see \S4). 
Column 12 summarizes the number of companions detected in the 
FGS observations (\S3), and the number of additional companions 
detected through spectroscopy or as wide visual binaries are 
given in columns 11 and 13, respectively (see \S4).  Column 14 lists
other commonly used names for the targets and a code to identify 
the LBV (or candidate LBV) stars. 

\placetable{tab1}  


\section{Companion Star Detection}                               

The detection of the signal of a stellar companion in the FGS scans depends 
primarily on the angular projected separation and magnitude difference. 
Each star in the FGS FOV produces a fringe pattern, and the 
observed scan will take the form 
\begin{equation}\label{eq1}
S(x)_{obs} = \sum_{i=1}^n f_i ~S(x-x_i)
\end{equation}
where each of $n$ stars contributes a flux fraction $f_i = \frac{F_i}{\sum F_j}$ 
and has a relative projected offset position $x_i$.  The function $S(\triangle x)$ 
represents the apparent fringe pattern produced by a single unresolved star. 
We show in Figure Set 1 the full collection of 251 merged scans of our targets 
(available in full in the electronic version of the paper), and the central 
two panels of these figures show the final scans along the orthogonal $x$ and $y$ 
axes.  A single star (cf.\ HD108, Fig.~1.1) shows a simple fringe oscillation 
pattern, while a fully resolved binary star (cf.\ HD73882, Fig.~1.135) shows two 
clearly separated fringe patterns.   In general, the relatively bright and 
widely separated companions are immediately detected upon inspection, 
but detection is more challenging with fainter companions or those cases of close 
companions where the fringe patterns largely overlap.  Our detection scheme
relies upon a comparison of the observed scans with those for a set of single 
stars that act as calibrator scans.  We first apply a set of detection tests
developed by \citet{2014AJ....147...40C}, and if a resolved component is
found, then we make a detailed fit of the observed scan with a selection of
calibrator scans.  Below we review the testing criteria and fitting procedure, 
and our results are summarized in Table 2 for resolved systems, Table 3 for 
partially resolved systems, and Table 4 for apparently single, unresolved systems. 

\placefigure{fig1} 

The detailed form of the fringe pattern $S(\triangle x)$ depends upon the 
color of the star and filter used \citep{2006AJ....132..836H} as well as the 
time of observation relative to that of a servicing mission or other adjustments of 
the instrument.  We selected the calibrators from a set of scans that appeared to be 
those of single stars from our program (see Table 4 below) and of red, low mass stars 
observed in programs GO-11943 and 11944.  These scans were subsequently checked for 
binary interlopers with the tests described below before establishing final 
lists of calibrator scans.  The scans were arranged into four categories based on 
filter (FND5 or F583W) and time of observation (before or after the final servicing 
mission on BY 2009.06), and they were ordered according to $B-V$ color.  
In most cases we relied upon all the available calibrators with colors within 
$\pm 0.5$ mag of the target's $B-V$ color (usually numbering between 6 and 50 cases). 

The first clues about the presence of a companion come from a visual inspection 
of the scans for multiple fringe patterns and from a measurement of the fringe 
amplitude dilution caused by the flux of the other star(s) (see eq.~\ref{eq1}).
The latter is measured by the $S$-curve peak-to-peak amplitude ratio 
(given as ``sppr'' in the central panels of Figure Set 1), which is the mean of
the ratio of observed to calibrator full amplitude among the set of selected
calibrators.  A value of sppr $<0.92$ is often an indication of the presence of
another flux source in the FGS FOV \citep{2014AJ....147...40C}. 

We tested for the presence of resolved companions using a cross-correlation 
function (CCF) method developed by \citet{2014AJ....147...40C}.  This is an 
iterative scheme that compares the CCF of a target with a calibrator scan 
to the CCF of the calibrator with itself.  The first step is to align and 
rescale the calibrator CCF with the main peak in the target CCF, and then 
this rescaled calibrator CCF is subtracted to search for residual peaks in the 
target CCF from companions.  The results of this first step (denoted RCCF for
``Residuals from the CCF subtraction'') are shown in the 
top two panels of Figure Set 1.  A vertical dashed line at the origin shows 
the position where the primary signal was removed.  Then, we sequentially identify 
any remaining CCF peaks that attain a strength $>4 \sigma$(CCF), where $\sigma$(CCF) 
is the standard deviation at that scan position among the collection of calibrator 
scan CCFs (shown as the light gray line in the upper panels of Figure Set 1). 
These peaks are also indicated by vertical dashed lines in Figure Set 1. 
Then, we use the scaling and offset parameters for each identified component to 
make a model composite scan, which is shown as a dashed line in the central panels
(often hidden within the line thickness of the observed scan plot).  
Finally, the difference between the observed and model scans is shown on an 
expanded amplitude scale in the lower panels of Figure Set 1, where the 
$\pm\sigma$(CCF) region is indicated by light gray shading.  

Note that all the CCF results shown in Figure Set 1 refer to a mean CCF 
derived from the CCFs of the target with each of the selected calibrator scans. 
Most of our targets are relatively bright and the merged scans have good 
S/N properties, so the main source of uncertainty in binary detection is 
related to how well the calibrator scans match the target scan.  Consequently, 
the detection criterion for a CCF peak is based on its strength relative to 
the scatter we find among the calibrator scans.  Simulation tests made 
by \citet{2012PhDT.........CN} indicate that our $>4 \sigma$(CCF) criterion 
will result in no more than a single accidental detection in a sample as large as ours. 
Indeed, there are potentially other plausible detection cases that can be 
made by inspection of the CCF plots in Figure Set 1, but for the purposes of this
paper, we generally include only those that meet this stringent requirement in order to 
avoid false detections. 

The CCF method yields ambiguous results for very close companions (with projected 
separations generally less than 20 mas) because in the first iteration the 
calibrator CCF will be matched to a position between the components where 
the composite CCF peaks.  In such a situation the residual CCF will show two 
comparable peaks around the origin.  Hence we require a second test to deal with 
close binaries that create blended fringe patterns.  \citet{2014AJ....147...40C}
showed that in such blended cases the difference between the observed and 
calibrator scans will have a functional shape proportional to the second derivative 
of the calibrator scan, 
\begin{equation}\label{eq2}
\begin{split}
S(x)_{\rm bin} - S(x)_{\rm cal} &
 = \frac{1}{1+r} S(-(\frac{r}{1+r}) \Delta x) + \frac{r}{1+r} S(+(\frac{1}{1+r}) \Delta x) - S(x) \\
& = \frac{1}{2} \frac{r}{(1+r)^2} (\Delta x)^2 S''(x) \equiv a S''(x) 
\end{split}
\end{equation}
where $\Delta x$ is the projected separation and $r=F_2/F_1$ is the flux ratio. 
This relation shows that the single parameter $a$ that can be derived from the 
blend is a function of both $r$ and $|\Delta x|$, so that these parameters cannot
be determined independently.  However, the relation also demonstrates that close
binaries can be detected by searching for those cases where the amplitude of the 
second derivative coefficient $a$ is large and positively valued.  We applied 
this second derivative test for detection by requiring $a > 4 \sigma(a)$, 
where $a$ and $\sigma(a)$ are the mean and standard deviation of fits of eq.~\ref{eq2}
from the set of selected calibrator scans.  Those cases that met this criterion 
are shown with a thick gray line portraying the fit in the lower panels of 
Figure Set 1 (cf.\ HD65087, Fig.~1.117). 

Once we had identified those resolved components with the CCF method, we then made 
a non-linear, least squares fit of the scans using the Levenberg-Marquardt algorithm 
with the IDL function {\it mpfitfun} \citep{2009ASPC..411..251M}. 
We did not make fits for the close blended scans because of the inherent 
ambiguity in the parameters in such cases (see eq.~\ref{eq2}). 
The binary or triple star fit was made of the positions and amplitudes of the fringe 
patterns for each component using a model of the form of eq.~\ref{eq1} but with 
independent parameters for the amplitude of each component.   
Starting values for each parameter were taken from the CCF results. 
The fits were made with each selected calibrator, and the final adopted values 
and uncertainties were estimated from the mean and standard deviation of the fitting 
parameters from the calibrator set. 

We found that it was preferable to have independent amplitude scaling parameters
for each component (rather than coefficients referenced to the flux of the primary 
as in eq.~\ref{eq1}) in order to deal effectively with the general scaling mismatch 
between the target and calibrator scans.  The magnitude differences were then 
obtained as $-2.5 \log F_i/F_1$ for each component.  In order to check our results, 
we compare in Figure~2 the derived magnitude differences (mean of $x$ and $y$-axis fits) 
with those obtained by {\it Hipparcos} \citep{1997ESASP1200.....P} 
for some of the mutually detected wide binaries.  The excellent agreement indicates that our 
magnitude estimates and their uncertainties are apparently reliable and free of 
systematic problems. 

\placefigure{fig2} 

There are generally four possible outcomes for binary detection along each axis:
(1) the fringe appearance is consistent with that of a single star, 
(2) the second derivative test indicates a blended component, 
(3) the fringes of a companion are resolved by the CCF test, or 
(4) a companion exists beyond the scan range but within the FGS FOV and 
causes a dilution of the fringe amplitude of the target (see eq.~\ref{eq1}).
If a system is triple, then the same set of outcomes is possible for the third component
(all dependent upon the orientation of the component in the sky relative to the scan axes). 
We attempted to decide upon these outcomes based upon an inter-comparison of the test
results between axes and the parameters of those known binary systems.  
The Appendix provides notes about those cases where the outcomes 
were ambiguous or problematical.   

The results for systems that were resolved along at least one axis are collected
in Table 2.  The entries are listed in order of increasing right ascension and by
date of observation where multiple observations were made.   
Columns 1 and 2 give the coordinates and name (same as in Table 1), and column 3
gives the discovery designation from the Washington Double Star (WDS) Catalog 
\citep{2001AJ....122.3466M}\footnote{http://ad.usno.navy.mil/wds/}. 
If the FGS observation is the first detection, then ``FGS'' is listed along with 
a component designation made following the nomenclature used in the WDS
\citep{2004RMxAC..21...83H}. 
Columns 4 and 5 give the date and filter for the observation. 
Columns 6, 7, and 8 give the position angle $\theta$, separation $\rho$, and 
magnitude difference $\triangle m$ determined by our non-linear, least squares 
fits of the scans.  In most cases the component is resolved in both axes. 
Then the position angle is determined from the projected axial separations 
and the telescope orientation on the sky 
(from the PA\_APER keyword in the observation header file),  
the separation is the square root of the 
sum of the squares of the projected axial separations, and 
the magnitude difference is the error weighted average of the $x$ and $y$ values. 
In other cases, the component is resolved on only one axis, but has a significant
second derivative coefficient $a$ for the other axis.  Then the absolute value of 
the close separation $|\Delta x|$ is derived using the flux ratio $r$ from the 
resolved axis result and the relation between $a$, $r$, and $|\Delta x|$ from 
eq.~\ref{eq2}.  This yields a reliable value for $\rho$, but there are two 
possible $\theta$ angles that correspond to the choice of $\pm|\Delta x|$. 
We list in Table 2 the $\theta$ estimate for $+|\Delta x|$ and the Appendix 
notes give the other possible $\theta$ value.  There are several cases 
where the companion is probably beyond the scan range along one axis, 
and for these there is no $\theta$ estimate and only a lower limit for $\rho$. 
Column 9 gives the number of the Figure Set 1 plot that corresponds to the 
observation, and column 10 provides codes for notes about the specific system. 

\placetable{tab2}  

Table 3 lists those cases where the second derivative test indicated 
the presence of a blended component along at least one axis (and 
the target is not included in Table 2).  Table 3 has the same format 
as Table 2, except for columns 6, 7, and 8 that are used differently. 
The second derivative coefficient $a$ depends on both flux ratio 
and separation (see eq.~\ref{eq2}), and we can set a minimum separation 
for a flux ratio $r=1$,
\begin{equation}\label{eq3}
\rho_{\rm min} = \sqrt{8} \sqrt{a_x +a_y} 
\end{equation}
where $a_x$ and $a_y$ are the positively valued, second derivative coefficients 
measured for the $x$ and $y$ scans, respectively.  This lower limit is
given in column 8 of Table 3.  If the flux ratio $r$ eventually
becomes known, then the actual separation will be given by 
\begin{equation}\label{eq4}
\rho = \rho_{\rm min} {{1+r}\over{2\sqrt{r}}}.
\end{equation}
There is a four-fold ambiguity in the derived position angle $\theta$
depending on the signs of $|\Delta x|$ and $|\Delta y|$.  Columns 6 and 7 give 
$^a\theta_1$ which is the ambiguous position angle for $(+|\Delta x|,+|\Delta y|)$
and $^a\theta_2$ which is the ambiguous position angle for $(+|\Delta x|,-|\Delta y|)$;
add $180^\circ$ to each of these to arrive at the remaining two possibilities. 
We can check on the validity of these estimates for the second derivative
detection of a close companion of HD37022C = $\theta^1$~Ori~C that is a binary with 
an orbit derived from long baseline interferometry \citep{2009A&A...497..195K}. 
\citet{2009A&A...497..195K} report a VLTI measurement at about the same time 
as the {\it HST} FGS observation with $\rho = 19.1$ mas and $\theta = 241^\circ$. 
If we adopt their optical flux ratio $r=0.30$, then eq.~\ref{eq4} and $\rho_{\rm min}$
yield estimates of $\rho = (17.9\pm 3.0)$ mas and $\theta = (247\pm19)^\circ$, 
in agreement with the contemporaneous VLTI measurement.  

\placetable{tab3}  

Table 4 lists the remaining systems for which we find no evidence of a companion. 
The format of Table 4 consists of the same first four and last two columns of Table 2.
Altogether, of our sample of 226 stars, we resolved 52 binary and 6 triple systems (Table 2), 
partially resolved 7 binaries (Table 3), leaving 161 stars unresolved (Table 4). 
Only 29 of the systems were known prior to this FGS survey.

\placetable{tab4}  

We show in Figure 3 the total separations and magnitude differences for all the 
components that we detected.  The partially resolved systems are plotted assuming
that the components have the same flux ($r=1$).  The solid line connecting 
diamond shaped symbols shows the expected faint limit for companion detection
by the CCF method, and the dotted line illustrates the expected limit for 
detection of close companions by the second derivative test (all for similar 
FGS scans from \citealt{2014AJ....147...40C}).  Our detections fall within the 
expected range for the most part, reaching as faint as $\triangle m = 5$ for widely 
separated binaries (but less for closer binaries).  The smallest separations 
we can detect are about 10 mas (Table 3).  For example, while we did detect the 
close binary HD37022C = $\theta^1$~Ori~C ($\rho = 19.1$ mas; Fig.~1.34), 
we failed to resolve the relatively bright companion of HD150136 (Fig.~1.196)
with a separation of 7 mas \citep{2013AnA...553A.131S,2013AnA...554L...4S}.  

\placefigure{fig3} 


\section{Companion Frequency}                                    

We found that 65 of 224 targets (omitting the subdwarf CD-45$^\circ$5058 = KS~292
and the K-giant star BD-3$^\circ$2178) or $29\%$ of the sample have a 
visual companion in the angular range from $0\farcs01$ to $1\farcs0$.   
This detection rate compares well with earlier surveys of massive stars 
in the Carina Association ($22\%$; \citealt{2004AJ....128..323N,2010AJ....139.2714N}) 
and in Cyg OB2 ($22\%$; \citealt{2014AJ....147...40C}).
We find 6 of the 13 LBV or candidate LBV stars to have companions, but four of
these are located in the LMC where source crowding is an issue, so we do 
not consider this high binary fraction to be unusual. 
However, in order to study the total multiplicity fraction, we must also consider
what is known about closer binaries (detected as spectroscopic binaries) and 
wider binaries (detected by speckle interferometry, adaptive optics, and other 
astrometric methods).  We have collected information on the binary companions of 
our sample through a literature review of the spectroscopic properties and 
a search through the WDS catalog for wider pairs. 
Furthermore, we have supplemented our sample of 224 stars with 81 others from 
the prior FGS surveys: 23 stars in the Carina association 
\citep{2004AJ....128..323N,2010AJ....139.2714N} 
(omitting HDE303308 which is already part of our main survey) and
58 stars from the Cyg~OB2 association \citep{2014AJ....147...40C}.
The information on these additional 81 stars is gathered at the bottom 
of Table~1 for the convenience of readers. 
Table 1, column 9 lists a code describing the spectroscopic status and 
column 10 gives a reference for the literature source.  Spectroscopic binaries
are identified with the code ``SB'' that is usually followed by the number of
spectral components observed (1 for a single-lined binary, 2 for a double-lined binary, 
and higher if additional components are known).  The code may include a suffix of 
``O'' for systems with orbital determinations, ``E'' for eclipsing or ellipsoidal systems, and 
``?'' for suspected systems (for example, for systems with a large radial velocity range 
but no orbit or those where double lines are reported).  A code of ``C'' indicates 
a star with apparent constant radial velocity.  Many of the targets are 
assigned a code of ``U'' for unknown status in cases where there are only a few or 
no radial velocity measurements.  The total number of probable spectroscopic companions 
(not including those detected in the FGS survey) is listed in column 11 of Table 1. 
Columns 12 and 13 give the numbers of visual companions found in our survey and 
from inspection of the WDS, respectively. 

We are also interested in the binary properties as a function of environment 
because stars ejected from their natal clusters may preferentially be single stars. 
Table 1, column 6 lists the name of the cluster or association of membership 
or the entry ``Field'' if no membership is known.  Most of these assignments 
come from earlier work by \citet{1978ApJS...38..309H}, 
\citet{1979A&AS...38..197M}, \citet{1982ApJ...263..777G}, 
and the cluster database WEBDA\footnote{http://www.univie.ac.at/webda/webda.html}.
We note that several of these ``clusters'' are in fact groups of only several 
luminous stars, but nevertheless, their existence shows that the target still resides 
among the stars where it was born.  \citet{2005AnA...437..247D} have shown 
that some so-called field stars are the brightest members of clusters with 
a host of fainter stars (e.g., HD52533, HD195592), and we suspect that many 
of the targets assigned to the field category in Table 1 may turn out to be
members of unrecognized clusters.   The runaway stars in the sample 
\citep{2009AJ....137.3358M} are indicated by an entry of ``yes'' in column 7 of Table 1.  


We caution that some of the wider, resolved companions  
may be field stars along the line of sight.  Furthermore, some of the 
targets reside in rich star clusters, and their companions may be cluster 
members that are not necessarily orbiting the primary target.  The probability 
of such a chance alignment may be estimated from the nearby surface density of 
stars with a magnitude less than that of the companion, $\Sigma(V<V_c)$. 
\citet{2006A&A...459..909C} show that the probability of finding a field 
star at a separation $\rho$ from the target is given by 
\begin{equation}\label{eq5}
p(\Sigma, \rho) = 1 - e^{-\pi \Sigma \rho^2}.
\end{equation}
We estimated $\Sigma$ in practice by collecting stellar $F$-magnitudes 
(covering the 579--642 nm range) in the region within a radius of $15\arcmin$ 
from the target that we extracted from the UCAC4 catalog \citep{2013AJ....145...44Z}.
We then formed cumulative distribution functions with magnitude for each set
and made a linear fit of the logarithm of cumulative star counts with magnitude
\citep{2014ApJ...785...47L}.  We used this fit to estimate $\Sigma$ for 
the magnitude of a given companion star, and then we estimated the probability
of chance alignment for the companion's projected separation $\rho$. 
Companions with a probability $p<0.01$ are good candidates for physically 
related objects.  

The companions detected in the FGS survey have small projected separations 
and are generally bright, so the probability of a chance alignment is 
much smaller than the $p=0.01$ criterion.  However, the situation is 
different for some of the more widely separated companions in the WDS sample.  
For example, there are seven companions listed in the WDS for HD190918, 
but only four of these meet the probability criterion.  This star is 
a member of the open cluster NGC6871, so it is possible that the 
remaining three companions are cluster members.  Long term proper motion 
investigations will be required to determine which of these companions 
are actually gravitationally bound to HD190918.   

Additional factors should be considered in assessing the status of the 
companions listed in the WDS.  For example, the runaway star HD34078 = AE~Aur
is listed with three companions in the WDS.  This is a surprising result
because this star was probably ejected from the Ori~OB1 association 
through an encounter between binary stars \citep{2004MNRAS.350..615G}, and 
the star is expected to be single at present.  A closer examination of the 
notes in the WDS indicates that the Aa,Ab companion is an artifact of 
adaptive optics imaging and that the AB companion is ``very doubtful''. 
Furthermore, according to UCAC4, the AC companion is probably 3.5 mag fainter 
than the magnitude listed in the WDS, so that its probability of chance alignment
is above the adopted criterion.  The tentative conclusion is that
HD34078 has no physically related companions, consistent with expectations. 
However, for the purposes of this work, we decided to retain all the
companions listed in the WDS, pending the further research that will be 
required to settle their true nature.  Thus, we caution that the 
companion numbers presented here from the WDS sample must be regarded 
as probable overestimates of the actual numbers of bound companions. 

Table~5 summarizes the numbers of companions according to their 
environmental parameter: cluster/association, field, or runaway groups.
We removed the four targets in the LMC from the total sample ($n=224+81-4=301$) 
because of crowding issues related to the large distance of the LMC. 
The companion numbers are first presented in section A for the resolved 
binaries in the FGS sample.  The number $n$(FGS) gives the number of 
targets with one or more detected companions in each environmental group. 
The next row gives the corresponding frequency of multiple systems 
($MF$ = number with any companion divided by the total number).
The uncertainty estimates are based upon the binomial statistical approach 
of \citet{2011PASA...28..128C} for a confidence interval of 
$c = 0.683$ (equivalent to $1 \sigma$), and they represent the average
of the almost equal lower and upper confidence limits. 
The third row reports the companion frequency 
($CF$ = number of companions divided by the number of targets).  
The uncertainties in this case were estimated by a bootstrap method
of random sampling of the data in the subsets (cf.\ \citealt{2010ApJS..190....1R}). 
The three rows in section B of Table~5 give the same values for the WDS sample. 
The estimates of $MF$ are similar for the two samples and the 
cluster/association and field stars, but the $MF$ estimate for the
runaway stars and the $CF$ estimates are all larger for the WDS sample. 
We suggest that this is due to overestimates of companion numbers in the WDS. 

\placetable{tab5}  

Section C of Table~5 lists the same $n$, $MF$, and $CF$ values
for the spectroscopic binaries in this sample.  The number $n$(SBO+E) counts
the number of targets with known orbital periods, i.e., those with 
spectroscopic orbits and/or eclipsing light curves.   The next three 
rows list the corresponding numbers for possible spectroscopic binaries
(with a status listing of ``SB1?'' or ``SB2?'' in Table 1), 
constant velocity stars, and stars with unknown spectroscopic binary 
properties, respectively.  Stars in the latter group were omitted in the 
calculation of $MF$ and $CF$ for the spectroscopic binaries.  The next 
four rows give the $MF$ and $CF$ estimates based upon two samples of the 
spectroscopic binaries, those with known period (SBO+E) and those 
known and suspected binaries (SBO+E+?).   

Section D of Table~5 gives the combined $MF$ and $CF$ estimates for 
two counting schemes.  The first combines the numbers of spectroscopic 
binaries with known period plus the numbers of FGS binaries.  
In this case we ignore any suspected spectroscopic binaries and all the 
WDS companions, so these statistics are noted as $MF$(min) and $CF$(min)
because they represent reliable minimum fractions.   
The second counting scheme sums all the known and suspected 
spectroscopic binaries, FGS companions, and WDS companions. 
These are representative of the observed maximum fractions, because 
they include some spectroscopic targets that may be velocity variable 
for reasons other than a binary companion and they include some unrelated 
companions from the WDS catalog.  

The high frequency of multiple systems among the SB category is similar to that 
found in recent spectroscopic surveys 
\citep{2012MNRAS.424.1925C,2012Sci...337..444S,2013A&A...550A.107S,2014ApJS..211...10S}, 
and our results confirm the trend that the ejected stars (runaway and some field stars)
have a lower frequency of multiple systems than stars still in their natal clusters. 
This trend is also seen among the FGS visual binaries, but it is probably absent
for the WDS sample because the bound companion numbers are overestimated for
our WDS sample (recall the case of the runaway star AE~Aur discussed previously). 
The relatively high frequency of multiple systems among the resolved binaries is also striking, 
and this verifies the importance of the more distant companions to the total 
numbers of companions \citep{1998AJ....115..821M,2009AJ....137.3358M,2012A&A...538A..74P}.  
The companion frequency is also very high among the cluster/association stars, reaching 
a value between 0.7 and 1.7 companions per target after combining the SB, FGS, and WDS samples. 

The numbers presented in Table~5 represent the properties of observed companions, 
and transforming these into the total numbers of multiple systems requires 
a careful consideration of observational selection effects and 
assumptions about the period and mass ratio distributions 
\citep{2012ApJ...751....4K,2013A&A...550A.107S}.
For example, the FGS survey is limited to companions brighter 
than $\triangle m = 5$, which corresponds approximately to $M_2/M_1 > 0.1$,
so we miss companions with a mass below a few solar masses.  
Such faint companions may be detected with adaptive optics (AO) imaging 
over a limited angular separation range \citep{2008AJ....136..554T}, 
but such AO observations are incomplete for our sample.  
Single high resolution measurements may also miss those systems that are 
close to a small separation conjunction phase at the time of observation. 
The spectroscopic binary numbers are based on observations with 
very diverse spectral resolution, wavelength coverage, and temporal 
cadence properties, and we suspect that many more binaries will 
be detected and/or verified in ongoing radial velocity investigations. 
Furthermore, the diversity of mass, age, and orbital periods in our sample 
may mix populations with differing binary properties \citep{2011A&A...528A.144K}. 
The binary statistics in Table~5 should therefore be regarded as the result 
of a convolution of the actual distributions with the observational 
selection effects that limit detections.


\section{Orbital Period Distribution}                            

We collected from the literature orbital periods for 83 of the SBO or SBE binary
systems listed in Table 1.   The visual binaries have much longer periods 
that are only beginning to be sampled, and there are published periods for 
only five visual binaries in our sample 
(HD37022, \citealt{2009A&A...497..195K};
 HD25639, \citealt{2007AstBu..62..352G};
 HD37468, \citealt{2008AJ....136..554T};
 HD47839, \citealt{2010NewA...15..302C};
 HD193322, \citealt{2011AJ....142...21T}).  
However, we may obtain an approximate orbital 
period for the visual binaries by considering their angular separation, 
distance, and probable mass.  The angular separation in the sky depends 
on orbital orientation and phase, and for circular orbits, we expect that
the projected separation generally underestimates the actual semimajor axis. 
On the other hand, many long period binaries have orbits with a large eccentricity, 
so that we observe them most of the time with a separation $(1+e)\times$ larger than 
the semimajor axis.  \citet{2006ApJ...652.1572B} made Monte-Carlo simulations of
the ratio of projected separation to semimajor axis for an ensemble of 
binaries with a commonly adopted eccentricity distribution $f(e)=2e$, 
and they found that this ratio has a value of $1.0\pm0.7$, where the uncertainty
represents the HWHM of the distribution (see their Fig.~9).  
Consequently, we estimated the semimajor axis $a$ for the visual binaries 
by $a \approx \rho d$, where $a$ is measured in AU, $\rho$ in arcsec, and $d$
in parsecs.  Table 1, column 8 lists the adopted distances for the targets, 
which were taken from WEBDA for cluster members and from \citet{2009MNRAS.400..518M}
for association stars.  Distances for the field stars were generally collected from 
spectroscopic parallaxes given by \citet{1980ApJ...242.1063G} or \citet{2012ApJS..199....8G}. 
If no distance estimate was found, then we calculated the spectroscopic
parallax ourselves using the magnitude, colors, and spectral classifications in Table~1
with intrinsic colors from \citet{1994MNRAS.270..229W}, a ratio of total-to-selective extinction
of $R=3.1$, and absolute magnitudes from \citet{1974MNRAS.166..203B} and \citet{2005A&A...436.1049M}.
We then estimated the orbital period $P$ using Kepler's Third Law and 
mass estimates for the primary from the spectral classification -- mass calibration
of \citet{2005A&A...436.1049M} (their Tables 4, 5, and 6). 
Note that we have ignored the need to adjust the period upwards because
the spectroscopic parallaxes probably underestimate the true distance 
(binaries are brighter than the primary alone), and likewise ignored a downwards period 
adjustment because the mass estimate is low (binaries are more massive than 
the single primary).  However, these changes are minor compared to the 
uncertainties inherent in our assumed equivalence of $a$ and $\rho$.  
Our final tally of orbital periods for visual binaries in our sample amounts to 
89 estimates for companions from the FGS detections plus 207 others for companions
from the WDS catalog. 

Our goal in this section is to determine the frequency of multiple systems $MF$ as 
function of the binary orbital period.  This requires a determination of the number
of targets in the sample for which our methods would probably find a binary over
a given period range.  Consequently, we need to consider the period range 
sensitivity for each method of binary detection.
A fortunate spectroscopic observer may discover a binary in a single measurement
of a double-lined system, but the determination of an orbital period for a spectroscopic
binary generally requires a significant effort of repeated observations.  Thus, 
the exploration space to determine a binary period grows with the 
number of observations and the duration between  
the first and last spectroscopic observations.  We extracted this observational 
duration from the papers cited in Table 1 for each of the targets with a spectroscopic
status different from ``U'' (unknown), and then we estimated the period detection 
range for each target as 1 day (smallest contact binary) to the full duration of 
observations.  We then constructed a logarithmic period grid using time in years 
and a bin size of 1 dex, and we determined the number of targets in each period bin
where the spectroscopic duration is sufficient to measure at least one binary orbital period.
This summation included cases where only a fraction of the $\log P$ bin range 
was covered by adding the ratio of the covered range to the full 1 dex bin size. 
Then the multiplicity fraction was calculated as the ratio of number 
of measured periods to the summed number of targets for which detection 
was possible within each $\log P$ bin. 
Note that our simple characterization of the period detection range fails to represent
the true complexity of the time series associated with the spectroscopic observations.
For example, a series of nightly observations made over one week plus a single 
observation made one month later would be taken at face value as suitable 
to detect periods up to one month, when, in fact, such a series is most
sensitive to periods of a week or less.  Thus, by using only the full duration 
of the observing sequence, we probably overestimate the detection efficiency 
at longer periods, and this may lead to a modest underestimate of $MF(\log P)$ 
at the longer orbital periods associated with the spectroscopic observations. 

We used a similar approach to find $MF(\log P)$ for the visual binaries
detected in the FGS survey and listed in the WDS catalog.  The period range
of detectability for these cases depends on the projected separation, 
distance, and stellar masses, and we used the distances from Table~1 
and masses from calibrations based upon spectral classification to 
determine $P$ from projected angular separation (in the same way as 
we did for the detected binaries).  The FGS scans are sensitive to 
binaries in the $0\farcs01$ to $1\farcs0$ range, while the WDS appears 
to list systems over a broader range of $\approx 0\farcs1$ to $\approx 100\arcsec$.
We adopted these angular ranges in setting the period range for binary detection 
for each target in our sample, and then we estimated the summed target number
and multiplicity fraction in each $\log P$ bin in the same way as for the 
spectroscopic sample.  Note that we took care not to double count those 
systems detected in the FGS survey that also appear in the WDS catalog. 

We show our resulting $MF(\log P)$ relation as a set of histograms in Figure 4.
The detected multiplicity fractions are shown individually for the 
SB, FGS, and WDS sets, and then the sum of these is shown as the 
final histogram (representing the total found from all methods).   
This summed distribution appears to be approximately flat, but we need to bear 
in mind a number of selection effects that may influence the appearance
of the distribution.  The low $\log P$ part of the distribution that is 
estimated from spectroscopic data is probably systematically low, 
because inclusion in the plot requires a significant observational 
effort, and we expect that a large fraction of the systems with a 
spectroscopic status of ``SB1?'' and ``SB2?'' in Table~1 will indeed 
turn out to be real short-period binaries.  Furthermore, it is likely
that observers may tend to favor short-period over long-period binaries,
because of the extended labor required to determine periods for 
the long-period systems.  On the other hand, it is relatively simple to 
estimate an approximate period for a visual binary from a single high angular 
resolution observation, and such observations are sensitive to relatively faint
and lower mass companions, so we might expect that the visual binary
$MF$ would tend to be relatively higher than the spectroscopic $MF$.  
We caution that the large number of companions found in the WDS may 
result partially from the inclusion of field stars or cluster members that may or 
may not be gravitationally bound to the target star.  This problem increases at the 
long end of the $\log P$ distribution (largest separation systems) where 
the estimated orbital periods become a significant fraction of the stellar 
lifetime.  If, for example, we replace the last two highly populated bins
in Figure~4 with the average in the shorter period bins, then the total
multiplicity fraction integrated over all period bins is 1.14,  
consistent will the idea that most massive stars have at least one companion. 

\placefigure{fig4} 

A number of investigators have explored the binary star period distribution,
with a particular emphasis on the shorter period systems ($P < 10$ d). 
\citet{2012ApJ...751....4K} developed a Monte-Carlo approach to sample the
intrinsic distributions of binary parameters in a way comparable to their
extensive spectroscopic observations of the massive stars in the Cyg~OB2 
association.  They used a power law distribution for orbital period of 
the form $f(\log P) \propto (\log P)^\beta$, and their experiments 
suggest $\beta=+0.2 \pm 0.4$, consistent with a flat 
distribution with $\beta=0$ (\"{O}pik's Law).  
On the other hand, \citet{2013A&A...550A.107S} 
used a similar Monte-Carlo method to fit spectroscopic results for a 
large sample of O-type stars in the Tarantula Nebula region of the LMC, 
and they find a best fit power law distribution with $\beta=-0.45 \pm 0.30$. 
Their result is consistent with that from an analysis of Milky Way eclipsing 
binaries by \citet{2013ApJ...778...95M}, who find $\beta = -0.4 \pm 0.3$. 
However, we caution that the distribution of shorter period systems 
may be more complicated and include a local maximum in numbers for 
periods in the range of 4 to 10 days
\citep{2010RMxAC..38...30B,2012ApJ...751....4K,2013A&A...550A.107S},
so that a multi-component model is more appropriate than a single power law 
\citep{2011IAUS..272..474S}.    Our results (Fig.~4) suggest that 
the distribution in $\log P$ is approximately flat when we consider the full 
range in orbital periods. 
 

\section{Conclusions}                                            

Our FGS survey has provided us with a new and uniform sample of high angular 
resolution observations to explore the multiple star properties of massive
stars in the projected separation range from $0\farcs01$ to $1\farcs0$ 
for companions brighter than $\triangle m = 5$ mag.  We used detection 
techniques developed by \citet{2014AJ....147...40C} to identify both faint 
companions and those close to the angular resolution limit.  In total, 
we detected 59 binary systems and 6 triple systems among our sample of 224 stars,
yielding a frequency of multiple systems of $29\%$.  Six of the 13 LBV or 
LBV candidates observed are found to have companions.  Many of the resolved 
binaries also have one component that is a spectroscopic binary, so our
results will help in the interpretation of their composite spectra. 
For example, all three of the bright stars BD+00$^\circ$1617 A,B,C
that line up in the center of the cluster Bochum~2 are resolved binaries, 
and two of these (B and C) are also short period spectroscopic 
binaries \citep{1999AnA...343..806M}, forming hierarchies like
those observed in the Orion Trapezium cluster \citep{2012ApJ...749..180C}.  
Although most of the resolved binaries are distant and the projected 
separations imply a large semimajor axis, we do find a number of relatively
nearby systems with close companions with probable orbital periods less 
than one century (including HD155913, HD158186, HDE229232, HDE303308, HD160529, 
HD164794  and HD195592).  These will be important targets for long term 
observation for orbital and mass determinations. 

We considered the binary star census of the complete sample 
(301 stars = 224 stars from this work less 4 LMC stars 
plus an additional 81 stars from earlier FGS studies)  
by collecting information from the literature on the numbers of close spectroscopic 
binaries and by searching the Washington Double Star Catalog for additional companions 
with angular separations mostly greater than one arcsec.  The number of companions 
was compared between the spectroscopic (SB) and resolved (FGS, WDS) samples
to determine the frequency of multiple systems and the companion frequency 
among stars residing in clusters and associations and in the field, and 
among runaway stars.  These statistics for the SB and FGS samples confirm the trend 
that stars close to their place of birth have relatively more companions, consistent 
with the idea that stars ejected from clusters are preferentially single objects. 
The number of wide companions in the WDS sample may be overestimated because
of the inclusion of cluster members and chance alignment cases rather than bound companions. 
The total number of companions per target among cluster and association stars
falls in the range from 0.7 to 1.7 depending upon the inclusion of suspected 
spectroscopic binaries and the WDS companions. 

We investigated the period distribution of the known binaries in this sample 
by collecting measured orbital periods for spectroscopic binaries and by 
estimating the periods for resolved binaries from their projected separation, 
distance, and probable mass.  We constructed a histogram of the multiplicity 
frequency as a function of $\log P$ by accounting for the probable range in 
detectable period for each target that was set by the duration of the observations 
for spectroscopic binaries and by the angular separation range associated with  
the FGS and WDS measurements for the visual binaries.  The resulting distribution 
is approximately flat over nine decades in $\log P$, consistent with \"{O}pik's Law.
However, there remain some significant observational selection effects 
that may eventually alter this conclusion.   Detailed spectroscopic and 
high angular resolution studies of massive stars in specific clusters with known 
distances will be particularly helpful in assessing the importance of 
such selection effects and determining the complete binary properties of a young 
massive star population (cf. \citealt{2012ApJ...751....4K,2013A&A...550A.107S}).  

\acknowledgments

We are grateful to Denise Taylor of STScI for her remarkably 
effective efforts that made these {\it HST} observations possible.
We also thank John Subasavage, Sergio Dieterich, and Adric Riedel
for their support of the FGS work at Georgia State University. 
Support for {\it HST} proposal numbers GO-11212, 11901, 11943, 
and 11944 was provided by NASA through
a grant from the Space Telescope Science Institute,
which is operated by the Association of Universities for Research
in Astronomy, Incorporated, under NASA contract NAS5-26555.
Institutional support has been provided from the GSU College
of Arts and Sciences and from the Research Program Enhancement
fund of the Board of Regents of the University System of Georgia,
administered through the GSU Office of the Vice President
for Research and Economic Development.  
JMA acknowledges support from [a] the Spanish
Government Ministerio de Econom{\'\i}a y Competitividad (MINECO) through grants
AYA2010-15\,081 and AYA2010-17\,631 and [b] the Consejer{\'\i}a de Educaci{\'o}n of
the Junta de Andaluc{\'\i}a through grant P08-TIC-4075.
AFJM is grateful to NSERC (Canada) and FQRNT 
(Quebec) for financial assistance.
NDR gratefully acknowledges his CRAQ (Quebec) fellowship.
This research has made use of the Washington Double Star Catalog 
maintained at the U.\ S.\ Naval Observatory and the WEBDA database, 
operated at the Institute for Astronomy of the University of Vienna.


Facilities: \facility{HST}


\appendix

\section{Notes on Individual Stars}


\noindent{\bf 024044.94+611656.1 = HD16429.}  
\citet{2003ApJ...595.1124M} found that the spectrum is a composite of 
an SB1 system and constant velocity component.  We assumed that one of 
these is the angularly resolved companion for counting purposes. 

\noindent{\bf 025107.97+602503.9 = HD17505.}  The companion is resolved on the $y$-axis only
and is far off-axis along the $x$-direction.  The results are consistent with the separation 
$2\farcs15$ and position angle $92\fdg7$ found by \citet{2010A&A...518A...1M}, 
although our estimated magnitude difference is slightly larger.  
\citet{2011ApJS..193...24S} obtained resolved spectra of both components and found 
that both are O-type stars.  Note that component A of this pair is itself
a spectroscopic triple star system \citep{2006ApJ...639.1069H}.

\noindent{\bf 040751.39+621948.4 = HD25639 = SZ Cam.}  Resolved on both axes for the first observation 
(at a position consistent with that found by 
\citealt{2007AstBu..62..339B}), but 
only resolved along the $y$-axis in the second observation.  We adopted
the magnitude difference from the first observation and the 
second derivative amplitude $a_x$ to estimate $|\triangle x|$ for
the second observation.  
\citet{2007AstBu..62..352G} show that the system 
consists of a short period eclipsing binary with a distant companion
that is probably also a binary (making the system a hierarchical quadruple).  
We assumed that the resolved component CHR 209 Ea,Eb is this second system 
for counting purposes. 

\noindent{\bf 051618.15+341844.3 = HD34078.}  We did not detect the close ($\rho = 0\farcs35$)
companion of AE~Aur discovered by \citet{2008AJ....136..554T} (TRN~17 Aa,Ab), 
which may have been an artifact of their adaptive optics observations (see \S4).

\noindent{\bf 051756.06-691603.9 = HDE269321.}  This close pair is resolved along the $y$-axis only 
in both of our closely spaced observations.  

\noindent{\bf 051814.36-691501.1 = HD35343 = S Dor.}  The companion is beyond the $x$-axis scan range 
in the second (short scan) observation.  

\noindent{\bf 053051.48-690258.6 = HDE269662.}  The companion is close, faint, and detected along the 
$y$-axis only in two closely spaced observations.  

\noindent{\bf 053522.90-052457.8 = HD37041A = $\theta ^2$ Ori A.} The CHR 249 Aa,Ab pair is clearly 
resolved in the first observation, but in the second short scan observation the companion is 
beyond the scan range in $x$ and is only partially resolved in the $y$ direction. 

\noindent{\bf 062715.78+145321.2 = HD45314.} 
\citet{1998AJ....115..821M} used speckle interferometry to resolve this 
target as a binary with a separation of $0\farcs054$ (named CHR 251 AB), but it 
was not resolved again in subsequent speckle observations 
\citep{2009AJ....137.3358M}.  It appears single in the FGS scans. 

\noindent{\bf 064548.70-071839.0 = ALS85.}  
This is a triple system where components B and C are comparable in 
brightness.  Consequently, the correspondence between the components observed 
in both axes is ambiguous.  Table 2 lists the result where the closer component
is assumed to be B in both cases.  If B has the larger projected separation 
in the $y$-axis scan, then the result for A,B is 
$\theta =  219\fdg87 \pm 0\fdg18$ and $\rho = 0\farcs3378 \pm 0\farcs0013$  
and the result for A,C is 
$\theta =  233\fdg29 \pm 0\fdg20$ and $\rho = 0\farcs3117 \pm 0\farcs0011$.

\noindent{\bf 071842.49-245715.8 = HD57061.}  $\tau$~CMa is a multiple system with 
two components revealed by the FGS observations.  The wider component was 
detected along both axes in the first observation, but only along the $x$-axis
in the second observation.  There is a low amplitude peak in the cross correlation 
function for the second observation near the expected projected separation 
($\triangle y \approx +0.19$) but it is below the adopted detection threshold. 
The system consists of a long period SB1 and a short period eclipsing system
\citep{1997AnA...327.1070V,1998Obs...118....7S},
and we assumed that these two correspond to the bright resolved pair FIN 313 Aa,Ab.
The WDS currently identifies Ab as the brighter of the two central objects, so 
we subtracted $180^\circ$ in position angle and changed the sign of $\triangle m$
to make our results consistent with the others in the WDS for the Aa,Ab pair. 
The wider component Ab,E appears in the WDS with a $180^\circ$ difference in position angle,
but a reassessment of AstraLux Lucky Imaging observations by \citet{2010A&A...518A...1M} 
indicates a placement consistent with the FGS results. 

\noindent{\bf 075220.28-262546.7 = HD64315.}  
This system was resolved as a binary by 
\citet{2009AJ....137.3358M} and named WSI 54 AB.  Recent observations by 
\citet{2012AJ....143...42H} agree with 
the position angle and separation estimated from the FGS observations (Table 2). 
However, speckle observations by 
\citet{2010AJ....139..743T} suggest that the 
system may consist of a triple in a linear configuration, and hence our binary
measurements may correspond to the center of light of the two companions.  The fit
of the $x$-axis scan with two components is marginal, but experiments with 
three component fits made little or no improvement, so we present the binary 
results in Table 2 for simplicity. 
\citet{2010ASPC..435..409L} present a spectroscopic
study and argue that the system consists of one SB2 system with a period of $P=2.71$~d 
plus one SBE system with a period of $1.018$~d.  We assumed that each of these 
correspond to components of the resolved binary for counting purposes. 



\noindent{\bf 081517.15-354414.6 = CD-35 4384.}  This is a triple system with an 
inner companion Ab detected by FGS.  It was difficult to rectify the low frequency 
trends in these long scans (particularly for the $x$-axis) and 
the magnitude difference for the wide pair Aa,B is taken from the $y$-axis result. 
Note that the actual uncertainty in magnitude difference may be larger than quoted in 
Table 2, because we do not account for spatial photometric sensitivity variations 
that become significant for widely separated systems.  

\noindent{\bf 081903.90-360844.9 = CD-35 4471.}  The companion was resolved along the $y$-axis only, but the 
second derivative test was nearly met for the $x$-axis result.  Thus, we estimated 
$|\triangle x|$ from the $y$-axis magnitude difference and second derivative amplitude $a_x$.
The result given in Table 2 corresponds to an assumed position at $+\triangle x$; for a 
projected position of $-\triangle x$, the position angle is $\theta = 146\fdg4 \pm 6\fdg1$. 

\noindent{\bf 084351.09-460346.5 = CD-45 4462.}  
The FGS scans reveal this as a triple system.  All three components 
appear in the $y$-axis scan, but the central pair is blended together in the $x$-axis scan. 
However, the second derivative amplitude is quite large for the central blend, so 
we estimated $|\triangle x|$ from the $y$-axis magnitude difference and second derivative amplitude $a_x$.
Table 2 lists the position angle of A,B for $+\triangle x$, and the position angle for 
$-\triangle x$ is $\theta = 20\fdg0 \pm 3\fdg3$.  All the magnitude differences are from 
the $y$-scan results. 

\noindent{\bf 085322.01-460208.8 = CD-45 4676.}  
The B companion is resolved along the $y$-axis and blended with the central 
fringe along the $x$-axis.   The second derivative test criterion is met in the latter case, 
so we estimated $|\triangle x|$ from the $y$-axis magnitude difference and second derivative amplitude $a_x$.
The position angle for $+\triangle x$ is given in Table 2, and that for $-\triangle x$ is
$\theta = 342\fdg31 \pm 0\fdg09$.


\noindent{\bf 090221.56-484154.4 = CD-48 4352.}  
This target appears as a triple in the $y$-axis scan and appears single in 
the $x$-axis scan.  However, the central fringe in the $x$-axis scan passes the second derivative
test, and we assume that the implied fringe broadening is due only to the closer and brighter 
B component (i.e., that the wider and fainter C component falls beyond the recorded $x$-axis scan). 
Then we estimated $|\triangle x|$ from the $y$-axis magnitude difference and second derivative amplitude $a_x$.
The $+|\triangle x|$ solution is used for the position angle in Table 2, and the result for $-|\triangle x|$ is
$\theta = 18\fdg6 \pm 2\fdg9$.

\noindent{\bf 100639.88-572533.1 = CPD-56 2853.}  The faint companion is resolved along the $y$-axis only. 
In this case the projected separation ($\triangle y = 0\farcs2$) is wide enough that 
we cannot say whether or not the the companion is blended or off-scan along the $x$-axis, 
and consequently we simply present a lower limit for the separation in Table 2. 

\noindent{\bf 104505.85-594006.4 = HDE303308.}  
This target was detected as a close binary in earlier FGS observations by 
\citet{2004AJ....128..323N,2010AJ....139.2714N} with $\theta = 122^\circ \pm 32^\circ$ 
and $\rho =0\farcs015 \pm 0\farcs002$
(resolved on the $y$-axis only).  We obtained two additional observations that do not 
resolve the system.  However, the second derivative test was suggestive of a companion 
(reaching a S/N ratio of 3.3 for the $y$-axis scan of the second observation, but still 
below our detection criterion of S/N $>4$).  Taking the second derivative amplitudes at their face values
yields the minimum separations and position angles given in Table 3.  Note that solution 
$^a\theta_2$ in the first observation is consistent with $^a\theta_1$ in the second observation.
The fact that three independent observations all yield similar binary parameters indicates that 
this system is probably a long period, wide binary.  The spectroscopic status is controversial. 
\citet{2012MNRAS.424.1925C} found the star to be radial velocity constant in ten observations. 
On the other hand, \citet{1991ApJS...75..869L} measured one very low radial velocity 
over an eight night run, consistent with a short period eccentric binary orbit.  
Consequently, we label the spectroscopic status as ``SB1?'' in Table~1. 

\noindent{\bf 164120.41-484546.6 = HD150136.}  
The companion resolved in $x$ only is consistent in position and magnitude difference
with the known A,B pair.  A companion with $\rho = 0\farcs0073$ detected in VLTI Amber 
observations by 
\citet{2013AnA...554L...4S} is too close to be resolved in the FGS data. 
\citet{2013AnA...553A.131S} discuss the orbits of the close binary and third star, and we include 
their period estimates in the spectroscopic category for Figure~4. 


\noindent{\bf 172912.93-313203.4 = HD158186.}  A companion is detected along the $y$-axis only.  
We adopt $\triangle x = 0$ in Table 2. 

\noindent{\bf 181512.97-202316.7 = HD167263.}  The close pair of 16~Sgr (CHR255 Aa,Ab) 
was observed in three previous speckle measurements with a position angle difference of 
$180^\circ$ from the FGS results, but this is not unexpected for stars of similar brightness.

\noindent{\bf 181805.90-121433.3 = HD167971.}  
\citet{2012MNRAS.423.2711D} resolved this system with the VLTI and argued that it 
has an orbital period $P> 20$~yr.  However, the separation was about 9 mas in 2008, 
which was too close for resolution with the somewhat noisier FGS scans we obtained. 
It is a hierarchical triple system with a close central binary. 

\noindent{\bf 182119.55-162226.1 = HD168625.}  
This target appears triple in the $x$-scans but double in the $y$-scan.  
It is not clear which of the two components in the $x$-scan corresponds to the single component in the $y$-scan, 
but we assumed that the component B with the smaller projected separation along the $x$-axis
corresponds to the resolved component along the $y$-axis (and that component C falls beyond
the range recorded for the $y$-scan).  The magnitude differences are taken from the $x$-axis data. 
The central fringe appeared somewhat asymmetrical in both $x$ and $y$ compared to those for
the calibrator stars.  Note that in the long scans made after 2009.1 (like this case) we 
often observe a weak feature at $\triangle x = -1\farcs2$ that has a systematic origin and 
should not be confused with a faint companion.  Only companion B is recorded along the $x$-axis in 
the second, short scan observation. Component B is probably the companion detected in 
VLT-NACO observations by \citet{2012ASPC..464..293M}. 

\noindent{\bf 200329.40+360130.5 = HD190429.}  Long scans were made to detect the signal of the wide B component. 
There are a few reports of a closer and fainter companion MCA 59 Aa,Ab at 
a separation of $\approx 0\farcs1$ (most recently by \citealt{1998AJ....115..821M}). 
However, this close companion is not detected in the FGS scans.  

\noindent{\bf 201806.99+404355.5 = HD193322A.}  
This is a remarkable multiple system that is the subject of a detailed 
study with the CHARA Array long baseline interferometer by \citet{2011AJ....142...21T}. 
The FGS observations resolve the Aa,Ab pair along the $y$-axis, but the pair is blended
in the $x$-axis scan.  A blend is indicated by the second derivative test and we used 
the $y$-scan magnitude difference and second derivative amplitude to find $|\triangle x|$.
The solution using $-|\triangle x|$ is listed in Table 2, and the separation and position angle 
estimates agree well with contemporaneous CHARA Array measurements (ten Brummelaar et al.\ 2011). 

\noindent{\bf 201851.71+381646.5 = HD193443A.} This system appears in the WDS with the brighter
component identified as B, so we added $180^\circ$ to the position angle and changed 
the sign of $\triangle m$ for consistency with the results in the WDS. 


\noindent{\bf 213857.62+572920.5 = HD206267.}  
This pair is resolved along the $y$-axis only, but the projected separation and 
magnitude difference are consistent with those for the known MIU 2 Aa,Ab system if the projected
separation is small along the $x$-axis.  The results in Table 2 assume $\triangle x = 0$.
The system is an hierarchical triple \citep{1995Obs...115..180S,1997ApJ...490..328B},  
and we assumed that the resolved companion is the third star identified in the spectrum. 



\bibliographystyle{apj}
\bibliography{apj-jour,paper}





\figsetstart
\figsetnum{1}
\figsettitle{FGS Scans and Binary Tests}
 
\figsetgrpstart
\figsetgrpnum{1.1}
\figsetgrptitle{f1}
\figsetplot{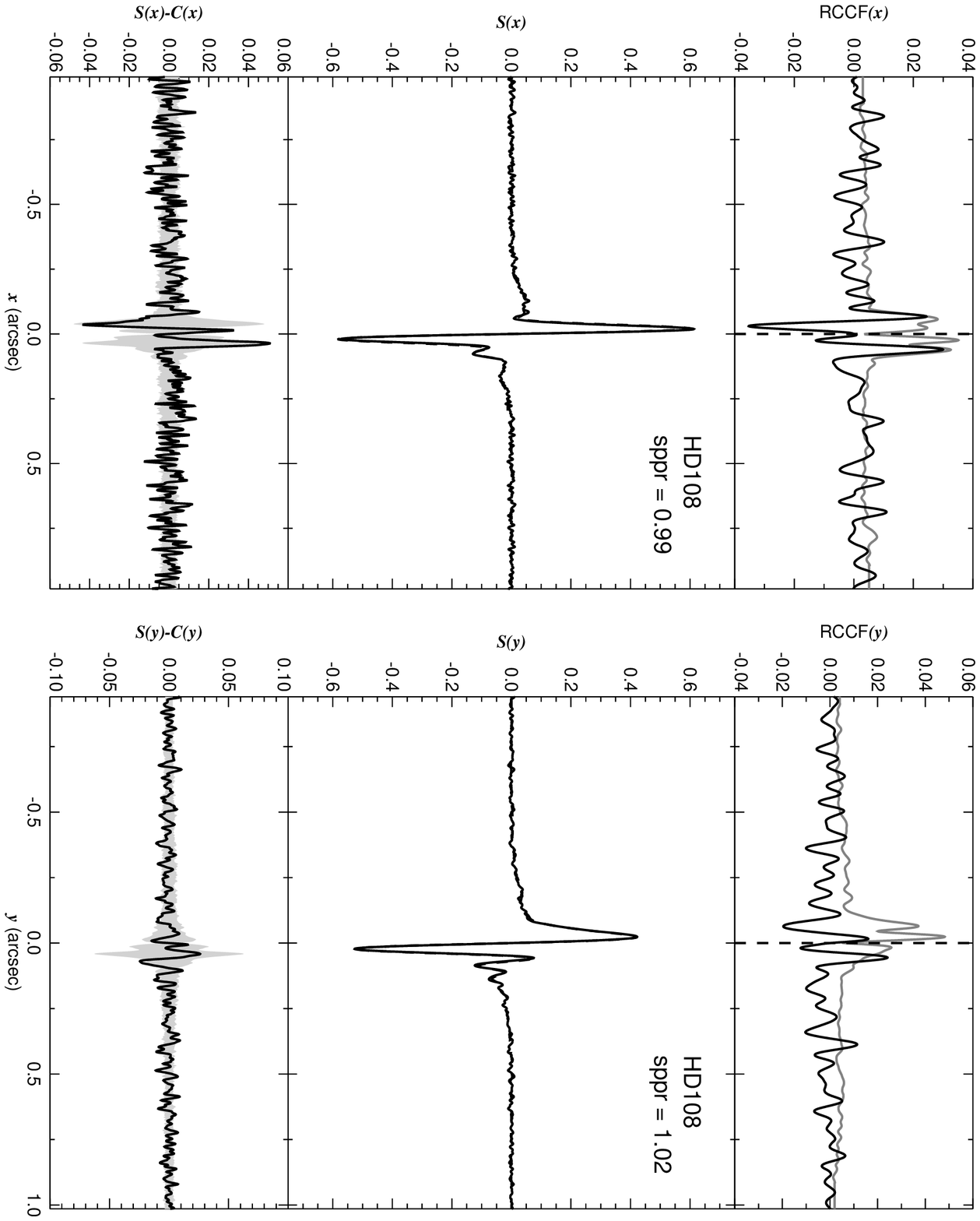}
\figsetgrpnote{The FGS scans and binary detection tests for target 
   000603.39$+$634046.8 = HD108
   obtained on BY 2008.5566.
   Figures 1.1 -- 1.251 are available in the online version of the Journal.}
\figsetgrpend
 
\figsetgrpstart
\figsetgrpnum{1.2}
\figsetgrptitle{f2}
\figsetplot{f1_2.eps}
\figsetgrpnote{The FGS scans and binary detection tests for target 
   001743.06$+$512559.1 = HD1337
   obtained on BY 2008.7090.}
\figsetgrpend
 
\figsetgrpstart
\figsetgrpnum{1.3}
\figsetgrptitle{f3}
\figsetplot{f1_3.eps}
\figsetgrpnote{The FGS scans and binary detection tests for target 
   014052.76$+$641023.1 = HD10125
   obtained on BY 2008.0775.}
\figsetgrpend
 
\figsetgrpstart
\figsetgrpnum{1.4}
\figsetgrptitle{f4}
\figsetplot{f1_4.eps}
\figsetgrpnote{The FGS scans and binary detection tests for target 
   022254.29$+$412847.7 = HD14633
   obtained on BY 2007.8425.}
\figsetgrpend

\figsetgrpstart
\figsetgrpnum{1.135}
\figsetgrptitle{f135}
\figsetplot{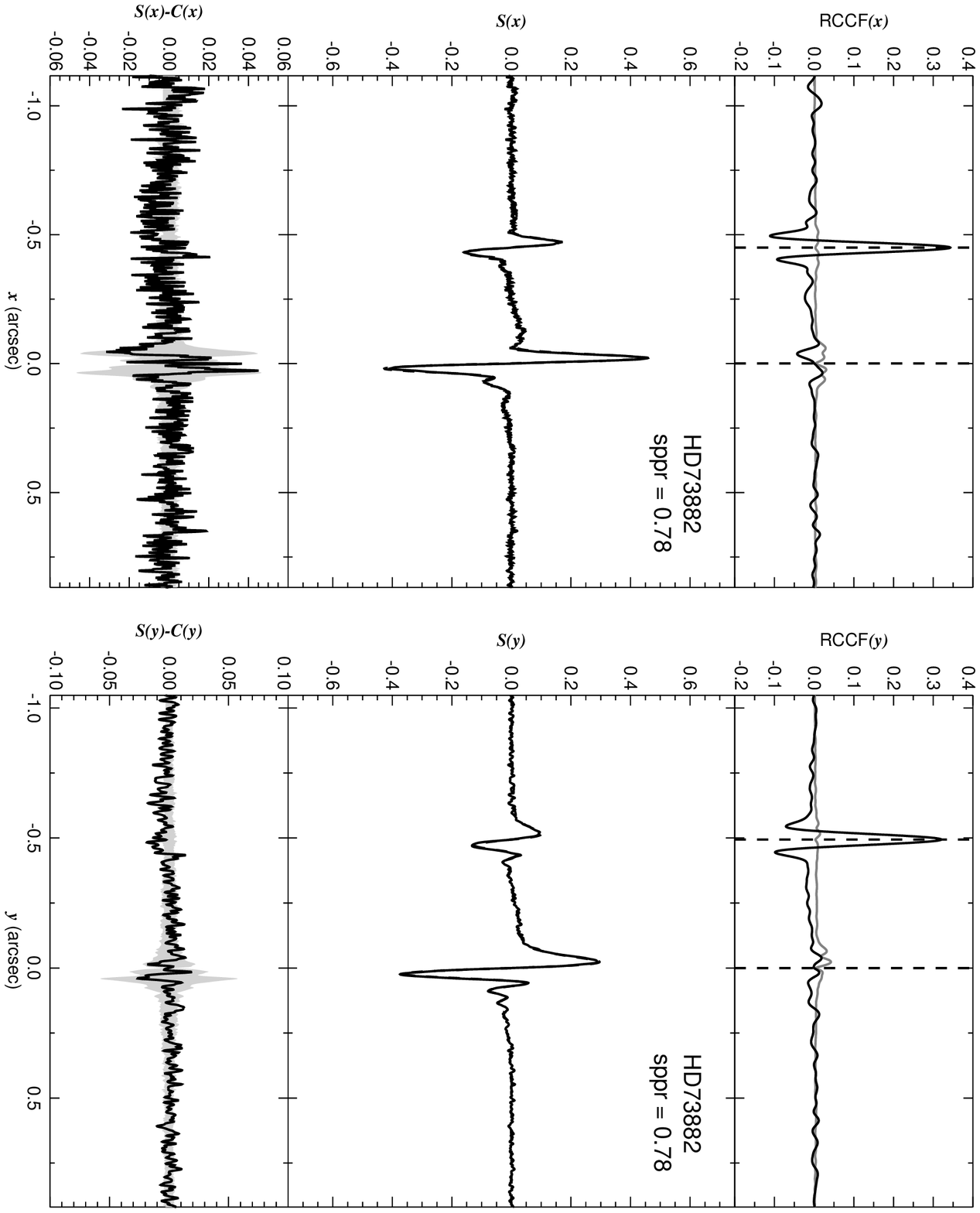}
\figsetgrpnote{The FGS scans and binary detection tests for target 
   083909.53$-$402509.3 = HD73882
   obtained on BY 2008.4061.}
\figsetgrpend

\figsetend

\clearpage
\begin{figure}
{\hspace*{-0.4cm}\includegraphics[angle=90,height=6.5cm]{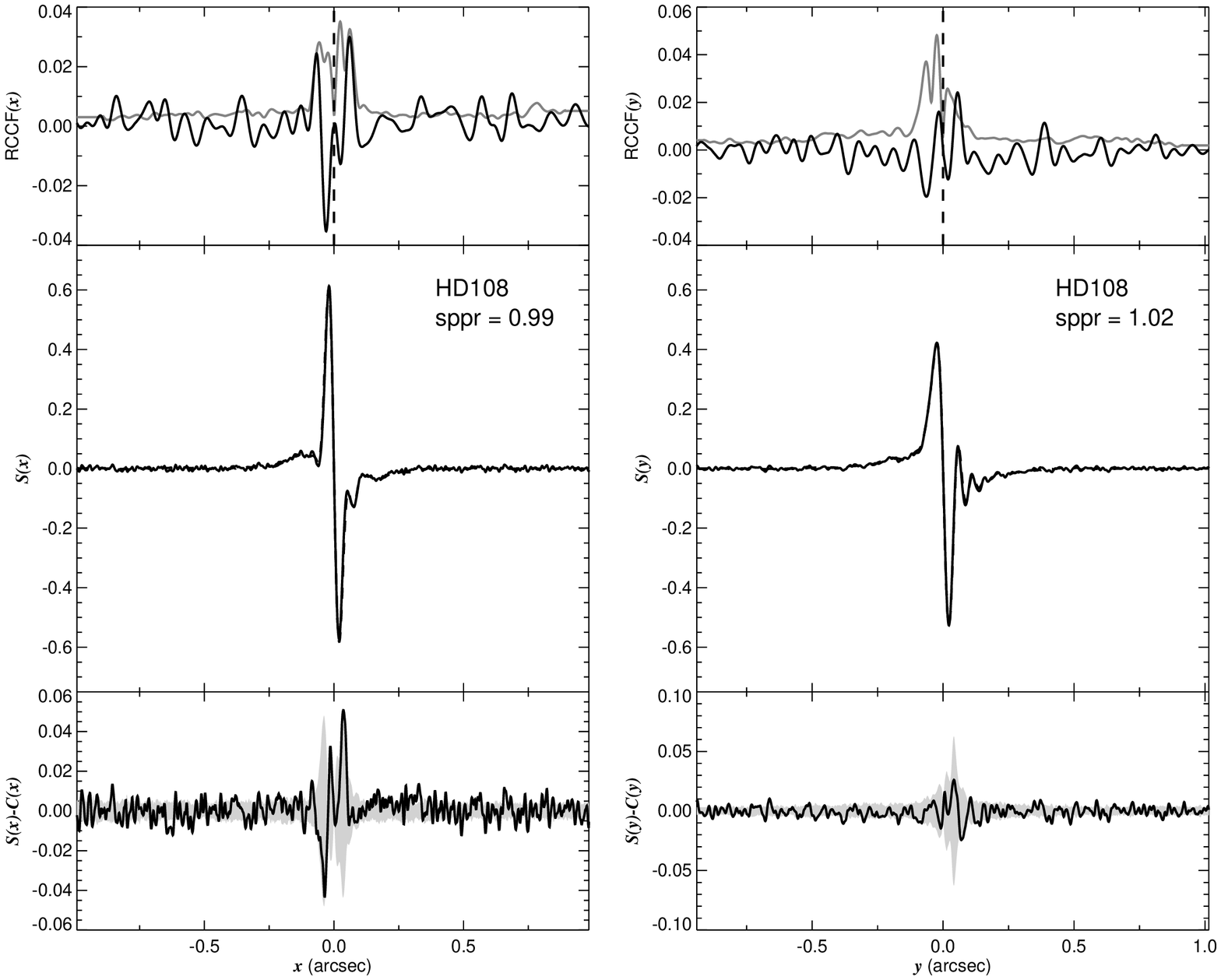}}
{\hspace*{-0.4cm}\includegraphics[angle=90,height=6.5cm]{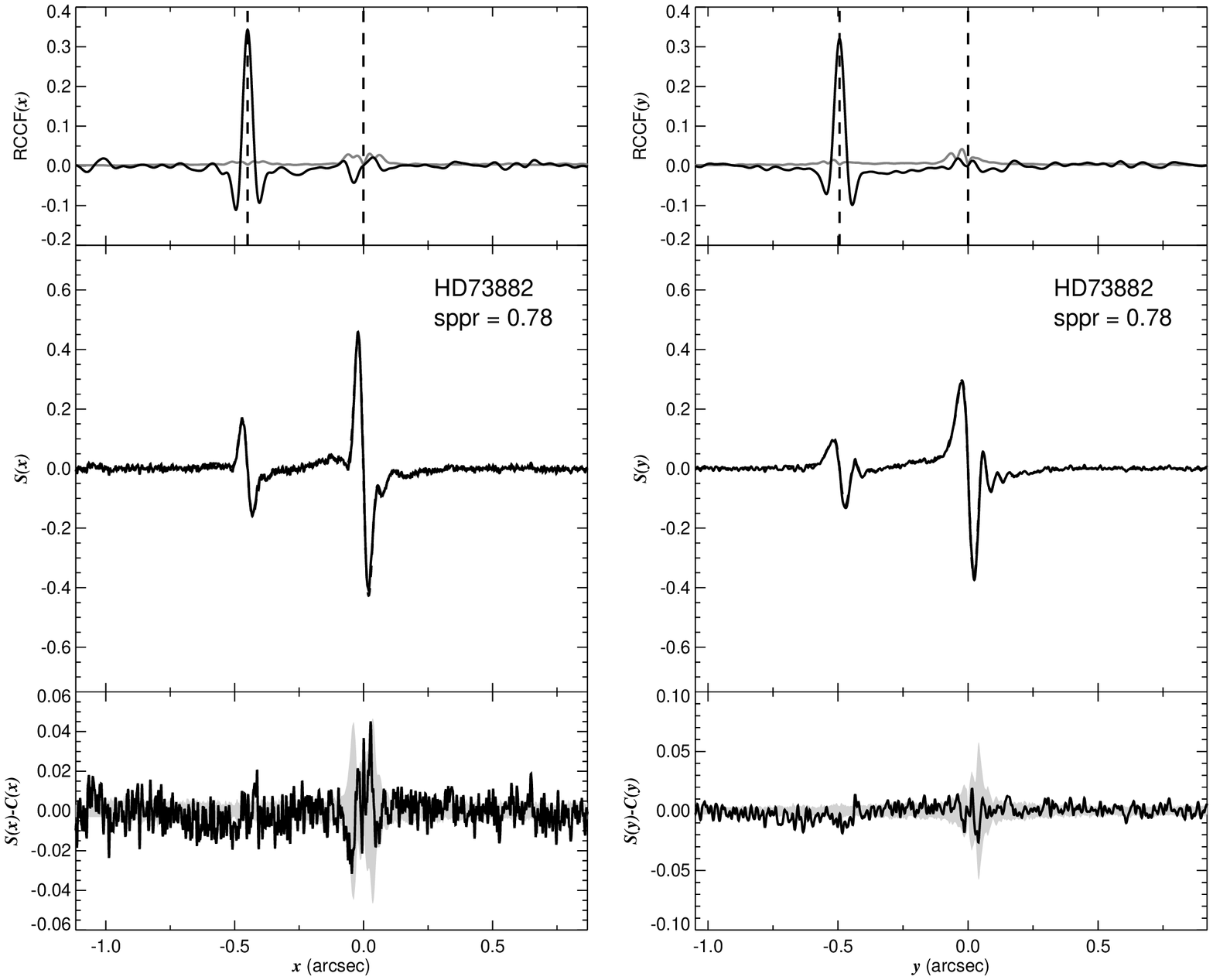}}
\caption{(A) The FGS scans and binary detection tests for the single star
   target 000603.39$+$634046.8 = HD108 obtained on BY 2008.5566.
   (B) The FGS scans and binary detection tests for the binary star target
   083909.53$-$402509.3 = HD73882 obtained on BY 2008.4061.
   Figures 1.1 -- 1.251 are available in the online version of the Journal.}
\end{figure}
 

\clearpage

\begin{figure}
\begin{center}
\plotone{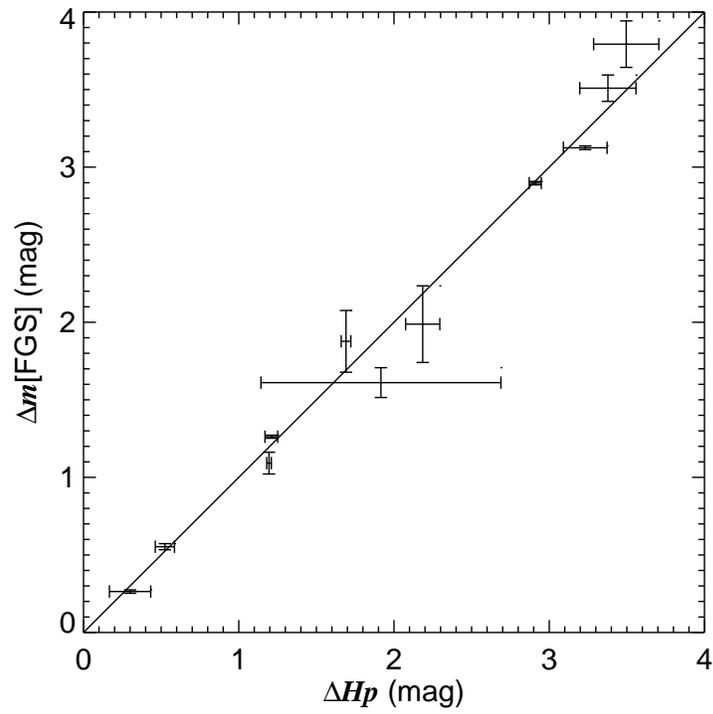}
\caption{A comparison of magnitude differences from {\it Hipparcos} and 
and FGS for pairs in common.  The estimates agree within uncertainties 
with the expected one-to-one relationship (shown as a solid line of 
slope unity).}
\label{fig2}
\end{center}
\end{figure}


\clearpage

\begin{figure}
\begin{center}
{\hspace*{1.0cm}\includegraphics[angle=90,height=12cm]{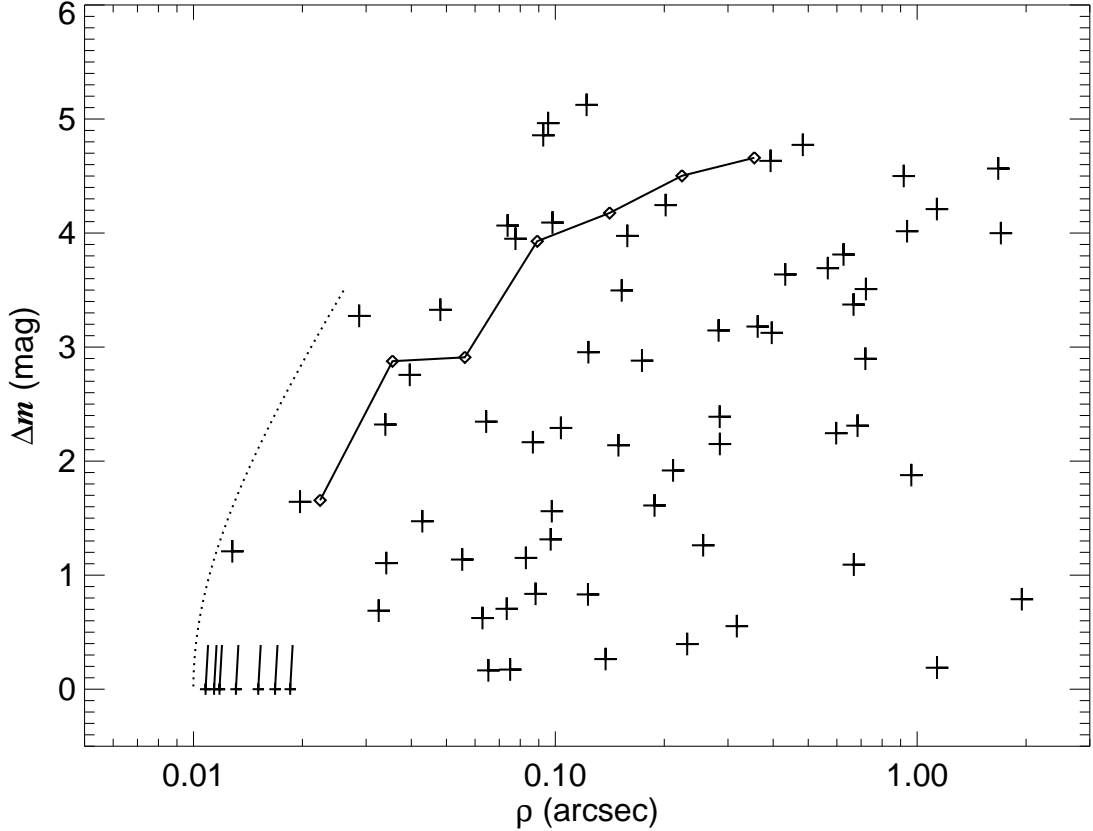}}
\caption{The fitted projected separation $\rho$ and magnitude difference 
$\triangle m$ for the resolved pairs (large plus signs) and the partially 
resolved pairs (small plus signs with line segments showing the displacement 
from $\triangle m=0.0$ to 0.4, i.e., for $F_2/F_1 =1.0 $ to 0.7).  The diamonds 
connected by a solid line represent the expected faint limits for detection by the 
cross correlation method and the dotted line shows the corresponding faint limit 
for detection by the second derivative test \citep{2014AJ....147...40C}.}
\label{fig3}
\end{center}
\end{figure}


\clearpage

\begin{figure}
\begin{center}
{\includegraphics[angle=0,height=14.0cm]{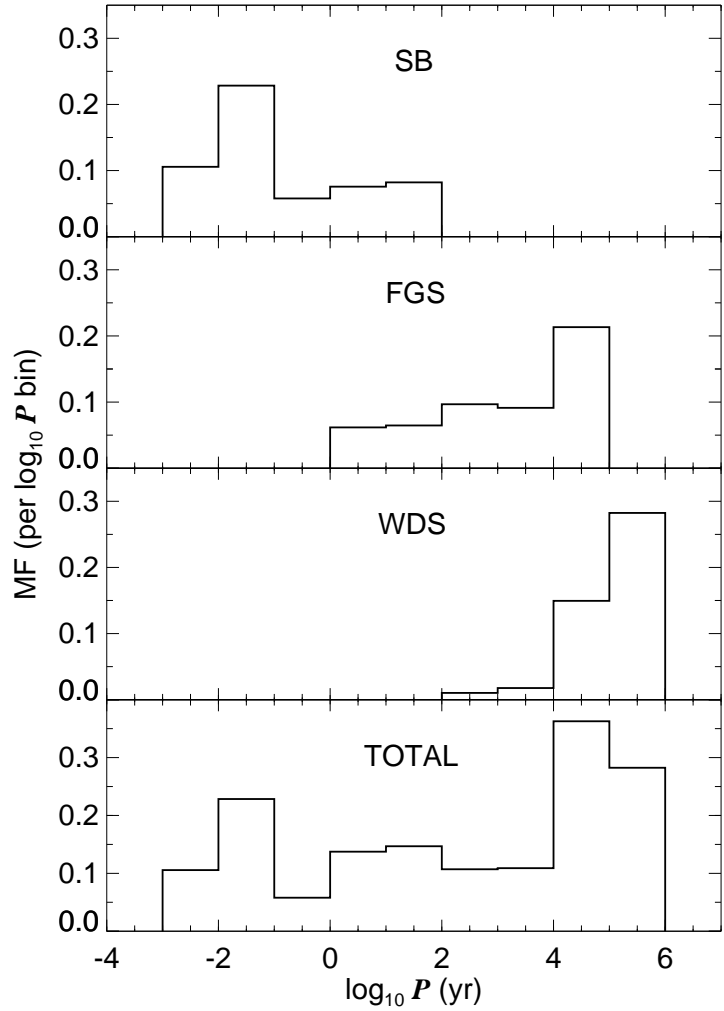}}
\caption{Histograms of the multiplicity fraction (MF) plotted as 
a function of orbital period.  From top to bottom successive panels
show the distributions for the spectroscopic binary (SB), 
Fine Guidance Sensor (FGS), Washington Double Star (WDS), and 
total samples, respectively.}
\label{fig4}
\end{center}
\end{figure}





\newpage
\begin{deluxetable}{lccccccccccccc}
\tabletypesize{\scriptsize}
\tablewidth{0pt}
\setlength{\tabcolsep}{0.02in} 
\rotate
\tablenum{1}
\tablecaption{Stellar Parameters\label{tab1}}
\tablehead{
\colhead{$(\alpha,\delta)$}  & 
\colhead{Star}               & 
\colhead{$V$}                & 
\colhead{$B-V$}              & 
\colhead{Spectral}           & 
\colhead{C/A/F}              & 
\colhead{Runaway}            & 
\colhead{$d$}                & 
\colhead{Spec.}              & 
\colhead{Spectroscopic}      & 
\colhead{$N$}                & 
\colhead{$N$}                & 
\colhead{$N$}                & 
\colhead{}                     
\\
\colhead{(J2000)}            & 
\colhead{Name}               & 
\colhead{(mag)}              & 
\colhead{(mag)}              & 
\colhead{Class.}             & 
\colhead{Category}           & 
\colhead{Status}             & 
\colhead{(kpc)}              & 
\colhead{Status}             & 
\colhead{Reference}          & 
\colhead{(SB)}               & 
\colhead{(FGS)}              & 
\colhead{(WDS)}              & 
\colhead{Notes}                
\\
\colhead{(1)} &
\colhead{(2)} &
\colhead{(3)} &
\colhead{(4)} &
\colhead{(5)} &
\colhead{(6)} &
\colhead{(7)} &
\colhead{(8)} &
\colhead{(9)} &
\colhead{(10)} &
\colhead{(11)} &
\colhead{(12)} &
\colhead{(13)} &
\colhead{(14)} 
}
\startdata
000603.39$+$634046.8 & HD108          & \phn7.39 & \phs0.17 &         O8 fpvar & 
            Cas OB5 &  no &  2.0 &     C & \citet{2001AnA...372..195N} & 
 0 &  0 &  2 &     \\
001743.06$+$512559.1 & HD1337         & \phn6.02 &  $-$0.05 &          O9.2 II & 
              Field & yes &  3.9 & SB2OE & \citet{1997Obs...117...37S} & 
 1 &  0 &  3 &  AO Cas   \\
014052.76$+$641023.1 & HD10125        & \phn8.22 & \phs0.31 &          O9.7 II & 
              Field &  no &  2.7 &  SB1? & \citet{2011AJ....142..146W} & 
 1 &  1 &  1 &     \\
022254.29$+$412847.7 & HD14633        & \phn7.46 &  $-$0.20 &          ON8.5 V & 
              Field & yes &  2.2 &  SB1O & \citet{2007ApJ...655..473M} & 
 1 &  1 &  1 &     \\
022759.81$+$523257.6 & HD15137        & \phn7.87 & \phs0.03 &     O9.5 II-IIIn & 
              Field & yes &  2.7 &  SB1O & \citet{2007ApJ...655..473M} & 
 1 &  0 &  0 &     \\
023249.42$+$612242.1 & HD15570        & \phn8.11 & \phs0.69 &            O4 If & 
            IC 1805 &  no &  1.9 &     C & \citet{2006ApJ...639.1069H} & 
 0 &  0 &  0 &     \\
024044.94$+$611656.1 & HD16429        & \phn7.67 & \phs0.62 & O9 II-III(n)Nwk  & 
            Cas OB6 &  no &  1.8 &  SB3O & \citet{2003ApJ...595.1124M} & 
 1 &  1 &  3 &     \\
024252.03$+$565416.5 & HD16691        & \phn8.70 & \phs0.48 &            O4 If & 
            Per OB1 &  no &  1.8 &     C & \citet{2009ApJ...704..964D} & 
 1 &  0 &  0 &     \\
025107.97$+$602503.9 & HD17505        & \phn7.07 & \phs0.40 &     O6.5 IIIn(f) & 
            IC 1848 &  no &  1.8 &  SB3O & \citet{2006ApJ...639.1069H} & 
 2 &  1 &  9 &     \\
025114.46$+$602309.8 & HD17520A       & \phn8.26 & \phs0.32 &            O8 Vz & 
            IC 1848 &  no &  1.8 &  SB2? & \citet{2006ApJ...639.1069H} & 
 1 &  1 & 13 &     \\
\enddata
\tablecomments{1 = LBV or LBV candidate;
 2 = FGS data from \citet{2004AJ....128..323N,2010AJ....139.2714N};
 3 = FGS data from \citet{2014AJ....147...40C} \\
 Table 1 is published in its entirety in the electronic 
edition of the {\it Astronomical Journal}.  A portion is shown here 
for guidance regarding its form and content.}
\end{deluxetable}



\newpage
\begin{deluxetable}{lllccccccc}
\tabletypesize{\scriptsize}
\tablewidth{0pt}
\setlength{\tabcolsep}{0.1in} 
\rotate
\tablenum{2}
\tablecaption{Resolved Companions\label{tab2}}
\tablehead{
\colhead{$(\alpha,\delta)$}  & 
\colhead{Star}               & 
\colhead{Discovery}          & 
\colhead{Date}               & 
\colhead{FGS}                & 
\colhead{$\theta$}           & 
\colhead{$\rho$}             & 
\colhead{$\triangle m$}      & 
\colhead{Fig.}               & 
\colhead{}                     
\\
\colhead{(J2000)}            & 
\colhead{Name}               & 
\colhead{Designation}        & 
\colhead{(BY)}               & 
\colhead{Filter}             & 
\colhead{(deg)}              & 
\colhead{(arcsec)}           & 
\colhead{(mag)}              & 
\colhead{1.$n$}              & 
\colhead{Notes}                
\\
\colhead{(1)} &
\colhead{(2)} &
\colhead{(3)} &
\colhead{(4)} &
\colhead{(5)} &
\colhead{(6)} &
\colhead{(7)} &
\colhead{(8)} &
\colhead{(9)} &
\colhead{(10)} 
}
\startdata
014052.76$+$641023.1 & HD10125        & HDS 221 AB    & 2008.0775 & F583W & 231.14$\pm$0.10 &  0.7216$\pm$0.0007 &  3.509$\pm$0.085 &   3 &     \\
022254.29$+$412847.7 & HD14633        & FGS 1 Aa,Ab   & 2007.8425 & F5ND  & 352.31$\pm$32.32&$>$0.0197$\pm$0.0111 & 1.643$\pm$1.083 &   4 & 2   \\
024044.94$+$611656.1 & HD16429        & CHR 208 Aa,Ab & 2007.6831 & F5ND  &  91.16$\pm$0.16 &  0.2849$\pm$0.0008 &  2.150$\pm$0.040 &   7 &     \\
025107.97$+$602503.9 & HD17505        & STF 306 AB    & 2008.5621 & F5ND  &  \nodata        &$>$0.2115$\pm$0.0007 & 1.918$\pm$0.054 &   9 & 1,5 \\
025114.46$+$602309.8 & HD17520A       & BU 1316 AB    & 2008.2139 & F583W & 298.83$\pm$0.08 &  0.3174$\pm$0.0008 &  0.553$\pm$0.020 &  10 &     \\
\enddata
\tablecomments{
1 = see Appendix;  
2 = resolved on the $x$-axis, unresolved on the $y$-axis, so 
the position angle and separation are estimated assuming $\triangle y = 0$;  
3 = resolved on the $y$-axis, unresolved on the $x$-axis, so 
the position angle and separation are estimated assuming $\triangle x = 0$;  
4 = resolved on the $x$-axis, off scan on the $y$-axis, so 
no position angle is listed and only a lower limit on the separation is given; 
5 = resolved on the $y$-axis, off scan on the $x$-axis, so 
no position angle is listed and only a lower limit on the separation is given;
6 = reassignment of bright star designation for consistency with WDS. \\
Table 2 is published in its entirety in the electronic
edition of the {\it Astronomical Journal}.  A portion is shown here
for guidance regarding its form and content.}
\end{deluxetable}



\newpage
\begin{deluxetable}{lllccccccc}
\tabletypesize{\scriptsize}
\tablewidth{0pt}
\setlength{\tabcolsep}{0.1in} 
\rotate
\tablenum{3}
\tablecaption{Partially Resolved Companions\label{tab3}}
\tablehead{
\colhead{$(\alpha,\delta)$}  & 
\colhead{Star}               & 
\colhead{Discovery}          & 
\colhead{Date}               & 
\colhead{FGS}                & 
\colhead{$^a\theta_1$}       & 
\colhead{$^a\theta_2$}       & 
\colhead{$\rho_{\rm min}$}   & 
\colhead{Fig.}               & 
\colhead{}                     
\\
\colhead{(J2000)}            & 
\colhead{Name}               & 
\colhead{Designation}        & 
\colhead{(BY)}               & 
\colhead{Filter}             & 
\colhead{(deg)}              & 
\colhead{(deg)}              & 
\colhead{(arcsec)}           & 
\colhead{1.$n$}              & 
\colhead{Notes}                
\\
\colhead{(1)} &
\colhead{(2)} &
\colhead{(3)} &
\colhead{(4)} &
\colhead{(5)} &
\colhead{(6)} &
\colhead{(7)} &
\colhead{(8)} &
\colhead{(9)} &
\colhead{(10)} 
}
\startdata
053516.47$-$052322.9 & HD37022C       & WGT 1 Ca,Cb & 2007.9029 & F5ND  & 247$\pm$19 &  35$\pm$19 & 0.0151$\pm$0.0025 &  34 &     \\
075557.13$-$283218.0 & HD65087        & FGS 33 AB   & 2008.8784 & F583W & 221$\pm$14 &  16$\pm$14 & 0.0168$\pm$0.0019 & 117 &     \\
104505.85$-$594006.4 & HDE303308      & NEL 5 Ha,Hb & 2008.5511 & F583W & 100$\pm$19 & 174$\pm$19 & 0.0096$\pm$0.0025 & 172 & 1   \\
104505.85$-$594006.4 & HDE303308      & NEL 5 Ha,Hb & 2008.8819 & F583W & 195$\pm$15 & 155$\pm$15 & 0.0114$\pm$0.0019 & 173 & 1   \\
110840.06$-$604251.7 & V432 Car       & FGS 34 AB   & 2008.9337 & F583W & 209$\pm$8  & 349$\pm$8  & 0.0118$\pm$0.0006 & 182 &     \\
110840.06$-$604251.7 & V432 Car       & FGS 34 AB   & 2008.9337 & F583W & 240$\pm$10 & 318$\pm$10 & 0.0075$\pm$0.0014 & 183 &     \\
174159.03$-$333013.7 & HD160529       & FGS 35 AB   & 2009.2643 & F5ND  & 191$\pm$17 &  11$\pm$17 & 0.0084$\pm$0.0037 & 208 &     \\
174159.03$-$333013.7 & HD160529       & FGS 35 AB   & 2009.2643 & F5ND  & 219$\pm$5  & 342$\pm$5  & 0.0131$\pm$0.0009 & 209 &     \\
180352.44$-$242138.6 & HD164794       & FGS 36 AB   & 2008.1920 & F5ND  & 246$\pm$13 & 292$\pm$13 & 0.0185$\pm$0.0019 & 213 &     \\
203034.97$+$441854.9 & HD195592       & FGS 37 AB   & 2008.5326 & F5ND  & 105$\pm$59 & 285$\pm$59 & 0.0108$\pm$0.0011 & 239 &         
\enddata
\tablecomments{
1 = see Appendix.}
\end{deluxetable}



\newpage
\begin{deluxetable}{llcccc}
\tabletypesize{\scriptsize}
\tablewidth{0pt}
\setlength{\tabcolsep}{0.1in} 
\tablenum{4}
\tablecaption{Unresolved Targets\label{tab4}}
\tablehead{
\colhead{$(\alpha,\delta)$}  & 
\colhead{Star}               & 
\colhead{Date}               & 
\colhead{FGS}                & 
\colhead{Fig.}               & 
\colhead{}                     
\\
\colhead{(J2000)}            & 
\colhead{Name}               & 
\colhead{(BY)}               & 
\colhead{Filter}             & 
\colhead{1.$n$}              & 
\colhead{Notes}                
\\
\colhead{(1)} &
\colhead{(2)} &
\colhead{(3)} &
\colhead{(4)} &
\colhead{(5)} & 
\colhead{(6)} 
}
\startdata
000603.39$+$634046.8 & HD108          & 2008.5566 & F5ND  &   1 &   \\
001743.06$+$512559.1 & HD1337         & 2008.7090 & F5ND  &   2 &   \\
022759.81$+$523257.6 & HD15137        & 2007.6777 & F5ND  &   5 &   \\
023249.42$+$612242.1 & HD15570        & 2007.6478 & F583W &   6 &   \\
024252.03$+$565416.5 & HD16691        & 2007.5274 & F583W &   8 &   \\
\enddata
\tablecomments{
1 = see Appendix.
Table 4 is published in its entirety in the electronic
edition of the {\it Astronomical Journal}.  A portion is shown here
for guidance regarding its form and content.}
\end{deluxetable}



\newpage
\begin{deluxetable}{lccc}
\tabletypesize{\small}
\tablewidth{0pt}
\tablenum{5}
\tablecaption{Frequency of Multiple Systems and Companion Frequency\label{tab5}}
\tablehead{
\colhead{Group}               & 
\colhead{Cluster/Association} & 
\colhead{Field}               & 
\colhead{Runaway}            \\ 
\colhead{(Number)}            & 
\colhead{(214)}               & 
\colhead{(58)}                & 
\colhead{(29)}                  
}
\startdata
\multicolumn{4}{c}{A. FGS Visual Binaries} \\
\tableline
$n$(FGS)      &            67 &             9 &             2 \\
$MF$(FGS)     & 0.31$\pm$0.03 & 0.16$\pm$0.05 & 0.07$\pm$0.05 \\
$CF$(FGS)     & 0.34$\pm$0.04 & 0.17$\pm$0.05 & 0.07$\pm$0.05 \\
\tableline
\multicolumn{4}{c}{B. WDS Visual Binaries} \\
\tableline
$n$(WDS)      &            61 &            10 &            10 \\
$MF$(WDS)     & 0.29$\pm$0.03 & 0.17$\pm$0.05 & 0.34$\pm$0.09 \\
$CF$(WDS)     & 0.84$\pm$0.14 & 0.22$\pm$0.07 & 0.52$\pm$0.16 \\
\tableline
\multicolumn{4}{c}{C. Spectroscopic Binaries} \\
\tableline
$n$(SBO+E)    &            68 &             5 &             5 \\
$n$(SB?)      &            28 &            14 &             3 \\
$n$(C)        &            65 &            14 &            21 \\
$n$(U)        &            53 &            25 &             0 \\
$MF$(SBO+E)   & 0.42$\pm$0.04 & 0.15$\pm$0.06 & 0.17$\pm$0.07 \\
$MF$(SBO+E+?) & 0.60$\pm$0.04 & 0.58$\pm$0.08 & 0.28$\pm$0.08 \\
$CF$(SBO+E)   & 0.51$\pm$0.05 & 0.15$\pm$0.06 & 0.17$\pm$0.07 \\
$CF$(SBO+E+?) & 0.68$\pm$0.05 & 0.58$\pm$0.09 & 0.28$\pm$0.08 \\
\tableline
\multicolumn{4}{c}{D. Any Companion} \\
\tableline
$MF$(min)     & 0.51$\pm$0.03 & 0.21$\pm$0.05 & 0.21$\pm$0.07 \\
$MF$(max)     & 0.69$\pm$0.03 & 0.50$\pm$0.06 & 0.48$\pm$0.09 \\
$CF$(min)     & 0.70$\pm$0.06 & 0.26$\pm$0.07 & 0.24$\pm$0.09 \\
$CF$(max)     & 1.67$\pm$0.17 & 0.72$\pm$0.11 & 0.86$\pm$0.20 \\
\enddata
\end{deluxetable}


\end{document}